\documentclass[a4paper]{aa}  
\usepackage{graphicx}
\usepackage{txfonts}
\graphicspath{{Figures/}}

\begin{document} 

\newcommand{\AASPP}{Astrononomy and Astrophysics Series, Ed. Pachart Publishing House Tucson}
\newcommand{\AJ}{AJ}
\newcommand{\APh}{Astroparticle Physics}
\newcommand{\ARAA}{ARA\&A}
\newcommand{\AaAS}{A\&As}
\newcommand{\AaA}{A\&A}
\newcommand{\AdSpR}{Advances in Space Research}
\newcommand{\ApJS}{Astrophys. J. Suppl. Ser.}
\newcommand{\ApJ}{ApJ}
\newcommand{\ApSS}{Astrophysics and Space Science}
\newcommand{\BAAS}{Bull. Am. Astron. Soc.}
\newcommand{\ExA}{Experimental Astronomy}
\newcommand{\IAUS}{IAU Symposium}
\newcommand{\MNRAS}{MNRAS}
\newcommand{\Natur}{Nature}
\newcommand{\PASAM}{Polish Academy of Science Arch Mech}
\newcommand{\PASJ}{Publications of the Astronomical Society of Japan}
\newcommand{\PASP}{Publications of the Astronomical Society of the Pacific}
\newcommand{\PhR}{Phys. Rep.}
\newcommand{\SSRv}{Space Sci. Rev.}
\newcommand{\Sci}{Science}
\newcommand{\JApA}{Journal of Astrophysics an Astronomy}
\newcommand{\PhRvD}{Phys. Rev. D}
\newcommand{\AmJPh}{American Journal of Physics}
\newcommand{\PhRv}{Physical Review}
\newcommand{\PhRvL}{Phys. Rev. Lett.}
\newcommand{\BaltA}{Baltic Astronomy}
\newcommand{\SvPhU}{Soviet Physics-Uspekhi}
\newcommand{\LNP}{Berlin Springer Verlag Lecture Notes in Physics}
\newcommand{\AN}{Astronomische Nachrichten}
\newcommand{\RPPh}{Reports of Progress in Physics}
\newcommand{\AuJPh}{Australian Journal of Physics}
\newcommand{\AARv}{Astronomy and Astrophysics Review}
\newcommand{\PhDT}{Ph.D. Thesis}
\newcommand{\CRASB}{Academie des Science Paris Comptes Rendus Serie B Sciences Physiques}
\newcommand{\PSS}{Planetary and Space Science}
\newcommand{\NCimC}{Nuovo Cimento C Geophysics Space Physics C}
\newcommand{\ITAS}{IEEE Transactions on Applied Superconductivity}
\newcommand{\JRASC}{Journal of the Royal Astronomical Society of Canada}
\newcommand{\Msngr}{The Messenger}

\def\setsymbol#1#2{\expandafter\def\csname #1\endcsname{#2}}
\def\getsymbol#1{\csname #1\endcsname}

\def\Planck{\textit{Planck}}

\def\Herschel{\textit{Herschel}}

\def\Spitzer{\textit{Spitzer}}

\def\HeJT{$^4$He-JT}

\def\allearlypapers{\nocite{planck2011-1.1, planck2011-1.3, planck2011-1.4, planck2011-1.5, planck2011-1.6, planck2011-1.7, planck2011-1.10, planck2011-1.10sup, planck2011-5.1a, planck2011-5.1b, planck2011-5.2a, planck2011-5.2b, planck2011-5.2c, planck2011-6.1, planck2011-6.2, planck2011-6.3a, planck2011-6.4a, planck2011-6.4b, planck2011-6.6, planck2011-7.0, planck2011-7.2, planck2011-7.3, planck2011-7.7a, planck2011-7.7b, planck2011-7.12, planck2011-7.13}}

\def\all2013resultspapers{\nocite{planck2013-p01, planck2013-p02, planck2013-p02a, planck2013-p02d, planck2013-p02b, planck2013-p03, planck2013-p03c, planck2013-p03f, planck2013-p03d, planck2013-p03e, planck2013-p01a, planck2013-p06, planck2013-p03a, planck2013-pip88, planck2013-p08, planck2013-p11, planck2013-p12, planck2013-p13, planck2013-p14, planck2013-p15, planck2013-p05b, planck2013-p17, planck2013-p09, planck2013-p09a, planck2013-p20, planck2013-p19, planck2013-pipaberration, planck2013-p05, planck2013-p05a, planck2013-pip56, planck2013-p06b}}

\newbox\tablebox    \newdimen\tablewidth
\def\leaderfil{\leaders\hbox to 5pt{\hss.\hss}\hfil}
%
%
\def\endPlancktable{\tablewidth=\columnwidth 
    $$\hss\copy\tablebox\hss$$
    \vskip-\lastskip\vskip -2pt}
\def\endPlancktablewide{\tablewidth=\textwidth 
    $$\hss\copy\tablebox\hss$$
    \vskip-\lastskip\vskip -2pt}
\def\tablenote#1 #2\par{\begingroup \parindent=0.8em
    \abovedisplayshortskip=0pt\belowdisplayshortskip=0pt
    \noindent
    $$\hss\vbox{\hsize\tablewidth \hangindent=\parindent \hangafter=1 \noindent
    \hbox to \parindent{$^#1$\hss}\strut#2\strut\par}\hss$$
    \endgroup}
\def\doubleline{\vskip 3pt\hrule \vskip 1.5pt \hrule \vskip 5pt}

%
\def\L2{\ifmmode L_2\else $L_2$\fi}
\def\dtt{\Delta T/T}
\def\DeltaT{\ifmmode \Delta T\else $\Delta T$\fi}
\def\deltat{\ifmmode \Delta t\else $\Delta t$\fi}
\def\fknee{\ifmmode f_{\rm knee}\else $f_{\rm knee}$\fi}
\def\Fmax{\ifmmode F_{\rm max}\else $F_{\rm max}$\fi}
\def\solar{\ifmmode{\rm M}_{\mathord\odot}\else${\rm M}_{\mathord\odot}$\fi}
\def\Msolar{\ifmmode{\rm M}_{\mathord\odot}\else${\rm M}_{\mathord\odot}$\fi}
\def\Lsolar{\ifmmode{\rm L}_{\mathord\odot}\else${\rm L}_{\mathord\odot}$\fi}
\def\mag{\sup{m}}
\def\inv{\ifmmode^{-1}\else$^{-1}$\fi}
\def\mo{\ifmmode^{-1}\else$^{-1}$\fi}
\def\sup#1{\ifmmode ^{\rm #1}\else $^{\rm #1}$\fi}
\def\expo#1{\ifmmode \times 10^{#1}\else $\times 10^{#1}$\fi}
\def\,{\thinspace}
\def\lsim{\mathrel{\raise .4ex\hbox{\rlap{$<$}\lower 1.2ex\hbox{$\sim$}}}}
\def\gsim{\mathrel{\raise .4ex\hbox{\rlap{$>$}\lower 1.2ex\hbox{$\sim$}}}}
\let\lea=\lsim
\let\gea=\gsim
\def\simprop{\mathrel{\raise .4ex\hbox{\rlap{$\propto$}\lower 1.2ex\hbox{$\sim$}}}}
\def\deg{\ifmmode^\circ\else$^\circ$\fi}
\def\pdeg{\ifmmode $\setbox0=\hbox{$^{\circ}$}\rlap{\hskip.11\wd0 .}$^{\circ}
          \else \setbox0=\hbox{$^{\circ}$}\rlap{\hskip.11\wd0 .}$^{\circ}$\fi}
\def\arcs{\ifmmode {^{\scriptstyle\prime\prime}}
          \else $^{\scriptstyle\prime\prime}$\fi}
\def\arcm{\ifmmode {^{\scriptstyle\prime}}
          \else $^{\scriptstyle\prime}$\fi}
\newdimen\sa  \newdimen\sb
\def\parcs{\sa=.07em \sb=.03em
     \ifmmode \hbox{\rlap{.}}^{\scriptstyle\prime\kern -\sb\prime}\hbox{\kern -\sa}
     \else \rlap{.}$^{\scriptstyle\prime\kern -\sb\prime}$\kern -\sa\fi}
\def\parcm{\sa=.08em \sb=.03em
     \ifmmode \hbox{\rlap{.}\kern\sa}^{\scriptstyle\prime}\hbox{\kern-\sb}
     \else \rlap{.}\kern\sa$^{\scriptstyle\prime}$\kern-\sb\fi}
\def\ra[#1 #2 #3.#4]{#1\sup{h}#2\sup{m}#3\sup{s}\llap.#4}
\def\dec[#1 #2 #3.#4]{#1\deg#2\arcm#3\arcs\llap.#4}
\def\deco[#1 #2 #3]{#1\deg#2\arcm#3\arcs}
\def\rra[#1 #2]{#1\sup{h}#2\sup{m}}
\def\page{\vfill\eject}
\def\dots{\relax\ifmmode \ldots\else $\ldots$\fi}
%
%
\def\WHzsr{\ifmmode $W\,Hz\mo\,sr\mo$\else W\,Hz\mo\,sr\mo\fi}
\def\mHz{\ifmmode $\,mHz$\else \,mHz\fi}
\def\GHz{\ifmmode $\,GHz$\else \,GHz\fi}
\def\mKs{\ifmmode $\,mK\,s$^{1/2}\else \,mK\,s$^{1/2}$\fi}
\def\muKs{\ifmmode \,\mu$K\,s$^{1/2}\else \,$\mu$K\,s$^{1/2}$\fi}
\def\muKRJs{\ifmmode \,\mu$K$_{\rm RJ}$\,s$^{1/2}\else \,$\mu$K$_{\rm RJ}$\,s$^{1/2}$\fi}
\def\muKHz{\ifmmode \,\mu$K\,Hz$^{-1/2}\else \,$\mu$K\,Hz$^{-1/2}$\fi}
\def\MJysr{\ifmmode \,$MJy\,sr\mo$\else \,MJy\,sr\mo\fi}
\def\MJysrmK{\ifmmode \,$MJy\,sr\mo$\,mK$_{\rm CMB}\mo\else \,MJy\,sr\mo\,mK$_{\rm CMB}\mo$\fi}
\def\microns{\ifmmode \,\mu$m$\else \,$\mu$m\fi}
\def\micron{\microns}
\def\muK{\ifmmode \,\mu$K$\else \,$\mu$\hbox{K}\fi}
\def\microK{\ifmmode \,\mu$K$\else \,$\mu$\hbox{K}\fi}
\def\muW{\ifmmode \,\mu$W$\else \,$\mu$\hbox{W}\fi}
\def\kms{\ifmmode $\,km\,s$^{-1}\else \,km\,s$^{-1}$\fi}
\def\kmsMpc{\ifmmode $\,\kms\,Mpc\mo$\else \,\kms\,Mpc\mo\fi}
%
%


\setsymbol{LFI:center:frequency:70GHz:units}{70.3\,GHz}
\setsymbol{LFI:center:frequency:44GHz:units}{44.1\,GHz}
\setsymbol{LFI:center:frequency:30GHz:units}{28.5\,GHz}

\setsymbol{LFI:center:frequency:70GHz}{70.3}
\setsymbol{LFI:center:frequency:44GHz}{44.1}
\setsymbol{LFI:center:frequency:30GHz}{28.5}

\setsymbol{LFI:center:frequency:LFI18:Rad:M:units}{71.7\GHz}
\setsymbol{LFI:center:frequency:LFI19:Rad:M:units}{67.5\GHz}
\setsymbol{LFI:center:frequency:LFI20:Rad:M:units}{69.2\GHz}
\setsymbol{LFI:center:frequency:LFI21:Rad:M:units}{70.4\GHz}
\setsymbol{LFI:center:frequency:LFI22:Rad:M:units}{71.5\GHz}
\setsymbol{LFI:center:frequency:LFI23:Rad:M:units}{70.8\GHz}
\setsymbol{LFI:center:frequency:LFI24:Rad:M:units}{44.4\GHz}
\setsymbol{LFI:center:frequency:LFI25:Rad:M:units}{44.0\GHz}
\setsymbol{LFI:center:frequency:LFI26:Rad:M:units}{43.9\GHz}
\setsymbol{LFI:center:frequency:LFI27:Rad:M:units}{28.3\GHz}
\setsymbol{LFI:center:frequency:LFI28:Rad:M:units}{28.8\GHz}
\setsymbol{LFI:center:frequency:LFI18:Rad:S:units}{70.1\GHz}
\setsymbol{LFI:center:frequency:LFI19:Rad:S:units}{69.6\GHz}
\setsymbol{LFI:center:frequency:LFI20:Rad:S:units}{69.5\GHz}
\setsymbol{LFI:center:frequency:LFI21:Rad:S:units}{69.5\GHz}
\setsymbol{LFI:center:frequency:LFI22:Rad:S:units}{72.8\GHz}
\setsymbol{LFI:center:frequency:LFI23:Rad:S:units}{71.3\GHz}
\setsymbol{LFI:center:frequency:LFI24:Rad:S:units}{44.1\GHz}
\setsymbol{LFI:center:frequency:LFI25:Rad:S:units}{44.1\GHz}
\setsymbol{LFI:center:frequency:LFI26:Rad:S:units}{44.1\GHz}
\setsymbol{LFI:center:frequency:LFI27:Rad:S:units}{28.5\GHz}
\setsymbol{LFI:center:frequency:LFI28:Rad:S:units}{28.2\GHz}

\setsymbol{LFI:center:frequency:LFI18:Rad:M}{71.7}
\setsymbol{LFI:center:frequency:LFI19:Rad:M}{67.5}
\setsymbol{LFI:center:frequency:LFI20:Rad:M}{69.2}
\setsymbol{LFI:center:frequency:LFI21:Rad:M}{70.4}
\setsymbol{LFI:center:frequency:LFI22:Rad:M}{71.5}
\setsymbol{LFI:center:frequency:LFI23:Rad:M}{70.8}
\setsymbol{LFI:center:frequency:LFI24:Rad:M}{44.4}
\setsymbol{LFI:center:frequency:LFI25:Rad:M}{44.0}
\setsymbol{LFI:center:frequency:LFI26:Rad:M}{43.9}
\setsymbol{LFI:center:frequency:LFI27:Rad:M}{28.3}
\setsymbol{LFI:center:frequency:LFI28:Rad:M}{28.8}
\setsymbol{LFI:center:frequency:LFI18:Rad:S}{70.1}
\setsymbol{LFI:center:frequency:LFI19:Rad:S}{69.6}
\setsymbol{LFI:center:frequency:LFI20:Rad:S}{69.5}
\setsymbol{LFI:center:frequency:LFI21:Rad:S}{69.5}
\setsymbol{LFI:center:frequency:LFI22:Rad:S}{72.8}
\setsymbol{LFI:center:frequency:LFI23:Rad:S}{71.3}
\setsymbol{LFI:center:frequency:LFI24:Rad:S}{44.1}
\setsymbol{LFI:center:frequency:LFI25:Rad:S}{44.1}
\setsymbol{LFI:center:frequency:LFI26:Rad:S}{44.1}
\setsymbol{LFI:center:frequency:LFI27:Rad:S}{28.5}
\setsymbol{LFI:center:frequency:LFI28:Rad:S}{28.2}


\setsymbol{LFI:white:noise:sensitivity:70GHz:units}{134.7\muKs}
\setsymbol{LFI:white:noise:sensitivity:44GHz:units}{164.7\muKs}
\setsymbol{LFI:white:noise:sensitivity:30GHz:units}{143.4\muKs}

\setsymbol{LFI:white:noise:sensitivity:70GHz}{134.7}
\setsymbol{LFI:white:noise:sensitivity:44GHz}{164.7}
\setsymbol{LFI:white:noise:sensitivity:30GHz}{143.4}


\setsymbol{LFI:white:noise:sensitivity:LFI18:Rad:M:units}{512.0\muKs}
\setsymbol{LFI:white:noise:sensitivity:LFI19:Rad:M:units}{581.4\muKs}
\setsymbol{LFI:white:noise:sensitivity:LFI20:Rad:M:units}{590.8\muKs}
\setsymbol{LFI:white:noise:sensitivity:LFI21:Rad:M:units}{455.2\muKs}
\setsymbol{LFI:white:noise:sensitivity:LFI22:Rad:M:units}{492.0\muKs}
\setsymbol{LFI:white:noise:sensitivity:LFI23:Rad:M:units}{507.7\muKs}
\setsymbol{LFI:white:noise:sensitivity:LFI24:Rad:M:units}{462.2\muKs}
\setsymbol{LFI:white:noise:sensitivity:LFI25:Rad:M:units}{413.6\muKs}
\setsymbol{LFI:white:noise:sensitivity:LFI26:Rad:M:units}{478.6\muKs}
\setsymbol{LFI:white:noise:sensitivity:LFI27:Rad:M:units}{277.7\muKs}
\setsymbol{LFI:white:noise:sensitivity:LFI28:Rad:M:units}{312.3\muKs}
\setsymbol{LFI:white:noise:sensitivity:LFI18:Rad:S:units}{465.7\muKs}
\setsymbol{LFI:white:noise:sensitivity:LFI19:Rad:S:units}{555.6\muKs}
\setsymbol{LFI:white:noise:sensitivity:LFI20:Rad:S:units}{623.2\muKs}
\setsymbol{LFI:white:noise:sensitivity:LFI21:Rad:S:units}{564.1\muKs}
\setsymbol{LFI:white:noise:sensitivity:LFI22:Rad:S:units}{534.4\muKs}
\setsymbol{LFI:white:noise:sensitivity:LFI23:Rad:S:units}{542.4\muKs}
\setsymbol{LFI:white:noise:sensitivity:LFI24:Rad:S:units}{399.2\muKs}
\setsymbol{LFI:white:noise:sensitivity:LFI25:Rad:S:units}{392.6\muKs}
\setsymbol{LFI:white:noise:sensitivity:LFI26:Rad:S:units}{418.6\muKs}
\setsymbol{LFI:white:noise:sensitivity:LFI27:Rad:S:units}{302.9\muKs}
\setsymbol{LFI:white:noise:sensitivity:LFI28:Rad:S:units}{285.3\muKs}

\setsymbol{LFI:white:noise:sensitivity:LFI18:Rad:M}{512.0}
\setsymbol{LFI:white:noise:sensitivity:LFI19:Rad:M}{581.4}
\setsymbol{LFI:white:noise:sensitivity:LFI20:Rad:M}{590.8}
\setsymbol{LFI:white:noise:sensitivity:LFI21:Rad:M}{455.2}
\setsymbol{LFI:white:noise:sensitivity:LFI22:Rad:M}{492.0}
\setsymbol{LFI:white:noise:sensitivity:LFI23:Rad:M}{507.7}
\setsymbol{LFI:white:noise:sensitivity:LFI24:Rad:M}{462.2}
\setsymbol{LFI:white:noise:sensitivity:LFI25:Rad:M}{413.6}
\setsymbol{LFI:white:noise:sensitivity:LFI26:Rad:M}{478.6}
\setsymbol{LFI:white:noise:sensitivity:LFI27:Rad:M}{277.7}
\setsymbol{LFI:white:noise:sensitivity:LFI28:Rad:M}{312.3}
\setsymbol{LFI:white:noise:sensitivity:LFI18:Rad:S}{465.7}
\setsymbol{LFI:white:noise:sensitivity:LFI19:Rad:S}{555.6}
\setsymbol{LFI:white:noise:sensitivity:LFI20:Rad:S}{623.2}
\setsymbol{LFI:white:noise:sensitivity:LFI21:Rad:S}{564.1}
\setsymbol{LFI:white:noise:sensitivity:LFI22:Rad:S}{534.4}
\setsymbol{LFI:white:noise:sensitivity:LFI23:Rad:S}{542.4}
\setsymbol{LFI:white:noise:sensitivity:LFI24:Rad:S}{399.2}
\setsymbol{LFI:white:noise:sensitivity:LFI25:Rad:S}{392.6}
\setsymbol{LFI:white:noise:sensitivity:LFI26:Rad:S}{418.6}
\setsymbol{LFI:white:noise:sensitivity:LFI27:Rad:S}{302.9}
\setsymbol{LFI:white:noise:sensitivity:LFI28:Rad:S}{285.3}


\setsymbol{LFI:knee:frequency:70GHz:units}{29.5\mHz}
\setsymbol{LFI:knee:frequency:44GHz:units}{56.2\mHz}
\setsymbol{LFI:knee:frequency:30GHz:units}{113.7\mHz}

\setsymbol{LFI:knee:frequency:70GHz}{29.5}
\setsymbol{LFI:knee:frequency:44GHz}{56.2}
\setsymbol{LFI:knee:frequency:30GHz}{113.7}

\setsymbol{LFI:knee:frequency:LFI18:Rad:M:units}{16.3\mHz}
\setsymbol{LFI:knee:frequency:LFI19:Rad:M:units}{15.1\mHz}
\setsymbol{LFI:knee:frequency:LFI20:Rad:M:units}{18.7\mHz}
\setsymbol{LFI:knee:frequency:LFI21:Rad:M:units}{37.2\mHz}
\setsymbol{LFI:knee:frequency:LFI22:Rad:M:units}{12.7\mHz}
\setsymbol{LFI:knee:frequency:LFI23:Rad:M:units}{34.6\mHz}
\setsymbol{LFI:knee:frequency:LFI24:Rad:M:units}{46.2\mHz}
\setsymbol{LFI:knee:frequency:LFI25:Rad:M:units}{24.9\mHz}
\setsymbol{LFI:knee:frequency:LFI26:Rad:M:units}{67.6\mHz}
\setsymbol{LFI:knee:frequency:LFI27:Rad:M:units}{187.4\mHz}
\setsymbol{LFI:knee:frequency:LFI28:Rad:M:units}{122.2\mHz}
\setsymbol{LFI:knee:frequency:LFI18:Rad:S:units}{17.7\mHz}
\setsymbol{LFI:knee:frequency:LFI19:Rad:S:units}{22.0\mHz}
\setsymbol{LFI:knee:frequency:LFI20:Rad:S:units}{8.7\mHz}
\setsymbol{LFI:knee:frequency:LFI21:Rad:S:units}{25.9\mHz}
\setsymbol{LFI:knee:frequency:LFI22:Rad:S:units}{15.8\mHz}
\setsymbol{LFI:knee:frequency:LFI23:Rad:S:units}{129.8\mHz}
\setsymbol{LFI:knee:frequency:LFI24:Rad:S:units}{100.9\mHz}
\setsymbol{LFI:knee:frequency:LFI25:Rad:S:units}{38.9\mHz}
\setsymbol{LFI:knee:frequency:LFI26:Rad:S:units}{58.9\mHz}
\setsymbol{LFI:knee:frequency:LFI27:Rad:S:units}{104.4\mHz}
\setsymbol{LFI:knee:frequency:LFI28:Rad:S:units}{40.7\mHz}

\setsymbol{LFI:knee:frequency:LFI18:Rad:M}{16.3}
\setsymbol{LFI:knee:frequency:LFI19:Rad:M}{15.1}
\setsymbol{LFI:knee:frequency:LFI20:Rad:M}{18.7}
\setsymbol{LFI:knee:frequency:LFI21:Rad:M}{37.2}
\setsymbol{LFI:knee:frequency:LFI22:Rad:M}{12.7}
\setsymbol{LFI:knee:frequency:LFI23:Rad:M}{34.6}
\setsymbol{LFI:knee:frequency:LFI24:Rad:M}{46.2}
\setsymbol{LFI:knee:frequency:LFI25:Rad:M}{24.9}
\setsymbol{LFI:knee:frequency:LFI26:Rad:M}{67.6}
\setsymbol{LFI:knee:frequency:LFI27:Rad:M}{187.4}
\setsymbol{LFI:knee:frequency:LFI28:Rad:M}{122.2}
\setsymbol{LFI:knee:frequency:LFI18:Rad:S}{17.7}
\setsymbol{LFI:knee:frequency:LFI19:Rad:S}{22.0}
\setsymbol{LFI:knee:frequency:LFI20:Rad:S}{8.7}
\setsymbol{LFI:knee:frequency:LFI21:Rad:S}{25.9}
\setsymbol{LFI:knee:frequency:LFI22:Rad:S}{15.8}
\setsymbol{LFI:knee:frequency:LFI23:Rad:S}{129.8}
\setsymbol{LFI:knee:frequency:LFI24:Rad:S}{100.9}
\setsymbol{LFI:knee:frequency:LFI25:Rad:S}{38.9}
\setsymbol{LFI:knee:frequency:LFI26:Rad:S}{58.9}
\setsymbol{LFI:knee:frequency:LFI27:Rad:S}{104.4}
\setsymbol{LFI:knee:frequency:LFI28:Rad:S}{40.7}


\setsymbol{LFI:slope:70GHz:units}{$-1.03$\mHz}
\setsymbol{LFI:slope:44GHz:units}{$-0.89$\mHz}
\setsymbol{LFI:slope:30GHz:units}{$-0.87$\mHz}

\setsymbol{LFI:slope:70GHz}{$-1.03$}
\setsymbol{LFI:slope:44GHz}{$-0.89$}
\setsymbol{LFI:slope:30GHz}{$-0.87$}

\setsymbol{LFI:slope:LFI18:Rad:M:units}{$-1.04$\mHz}
\setsymbol{LFI:slope:LFI19:Rad:M:units}{$-1.09$\mHz}
\setsymbol{LFI:slope:LFI20:Rad:M:units}{$-0.69$\mHz}
\setsymbol{LFI:slope:LFI21:Rad:M:units}{$-1.56$\mHz}
\setsymbol{LFI:slope:LFI22:Rad:M:units}{$-1.01$\mHz}
\setsymbol{LFI:slope:LFI23:Rad:M:units}{$-0.96$\mHz}
\setsymbol{LFI:slope:LFI24:Rad:M:units}{$-0.83$\mHz}
\setsymbol{LFI:slope:LFI25:Rad:M:units}{$-0.91$\mHz}
\setsymbol{LFI:slope:LFI26:Rad:M:units}{$-0.95$\mHz}
\setsymbol{LFI:slope:LFI27:Rad:M:units}{$-0.87$\mHz}
\setsymbol{LFI:slope:LFI28:Rad:M:units}{$-0.88$\mHz}
\setsymbol{LFI:slope:LFI18:Rad:S:units}{$-1.15$\mHz}
\setsymbol{LFI:slope:LFI19:Rad:S:units}{$-1.00$\mHz}
\setsymbol{LFI:slope:LFI20:Rad:S:units}{$-0.95$\mHz}
\setsymbol{LFI:slope:LFI21:Rad:S:units}{$-0.92$\mHz}
\setsymbol{LFI:slope:LFI22:Rad:S:units}{$-1.01$\mHz}
\setsymbol{LFI:slope:LFI23:Rad:S:units}{$-0.95$\mHz}
\setsymbol{LFI:slope:LFI24:Rad:S:units}{$-0.73$\mHz}
\setsymbol{LFI:slope:LFI25:Rad:S:units}{$-1.16$\mHz}
\setsymbol{LFI:slope:LFI26:Rad:S:units}{$-0.79$\mHz}
\setsymbol{LFI:slope:LFI27:Rad:S:units}{$-0.82$\mHz}
\setsymbol{LFI:slope:LFI28:Rad:S:units}{$-0.91$\mHz}

\setsymbol{LFI:slope:LFI18:Rad:M}{$-1.04$}
\setsymbol{LFI:slope:LFI19:Rad:M}{$-1.09$}
\setsymbol{LFI:slope:LFI20:Rad:M}{$-0.69$}
\setsymbol{LFI:slope:LFI21:Rad:M}{$-1.56$}
\setsymbol{LFI:slope:LFI22:Rad:M}{$-1.01$}
\setsymbol{LFI:slope:LFI23:Rad:M}{$-0.96$}
\setsymbol{LFI:slope:LFI24:Rad:M}{$-0.83$}
\setsymbol{LFI:slope:LFI25:Rad:M}{$-0.91$}
\setsymbol{LFI:slope:LFI26:Rad:M}{$-0.95$}
\setsymbol{LFI:slope:LFI27:Rad:M}{$-0.87$}
\setsymbol{LFI:slope:LFI28:Rad:M}{$-0.88$}
\setsymbol{LFI:slope:LFI18:Rad:S}{$-1.15$}
\setsymbol{LFI:slope:LFI19:Rad:S}{$-1.00$}
\setsymbol{LFI:slope:LFI20:Rad:S}{$-0.95$}
\setsymbol{LFI:slope:LFI21:Rad:S}{$-0.92$}
\setsymbol{LFI:slope:LFI22:Rad:S}{$-1.01$}
\setsymbol{LFI:slope:LFI23:Rad:S}{$-0.95$}
\setsymbol{LFI:slope:LFI24:Rad:S}{$-0.73$}
\setsymbol{LFI:slope:LFI25:Rad:S}{$-1.16$}
\setsymbol{LFI:slope:LFI26:Rad:S}{$-0.79$}
\setsymbol{LFI:slope:LFI27:Rad:S}{$-0.82$}
\setsymbol{LFI:slope:LFI28:Rad:S}{$-0.91$}


\setsymbol{LFI:FWHM:70GHz:units}{13\parcm01}
\setsymbol{LFI:FWHM:44GHz:units}{27\parcm92}
\setsymbol{LFI:FWHM:30GHz:units}{32\parcm65}

\setsymbol{LFI:FWHM:70GHz}{13.01}
\setsymbol{LFI:FWHM:44GHz}{27.92}
\setsymbol{LFI:FWHM:30GHz}{32.65}

\setsymbol{LFI:FWHM:LFI18:units}{13\parcm39}
\setsymbol{LFI:FWHM:LFI19:units}{13\parcm01}
\setsymbol{LFI:FWHM:LFI20:units}{12\parcm75}
\setsymbol{LFI:FWHM:LFI21:units}{12\parcm74}
\setsymbol{LFI:FWHM:LFI22:units}{12\parcm87}
\setsymbol{LFI:FWHM:LFI23:units}{13\parcm27}
\setsymbol{LFI:FWHM:LFI24:units}{22\parcm98}
\setsymbol{LFI:FWHM:LFI25:units}{30\parcm46}
\setsymbol{LFI:FWHM:LFI26:units}{30\parcm31}
\setsymbol{LFI:FWHM:LFI27:units}{32\parcm65}
\setsymbol{LFI:FWHM:LFI28:units}{32\parcm66}

\setsymbol{LFI:FWHM:LFI18}{13.39}
\setsymbol{LFI:FWHM:LFI19}{13.01}
\setsymbol{LFI:FWHM:LFI20}{12.75}
\setsymbol{LFI:FWHM:LFI21}{12.74}
\setsymbol{LFI:FWHM:LFI22}{12.87}
\setsymbol{LFI:FWHM:LFI23}{13.27}
\setsymbol{LFI:FWHM:LFI24}{22.98}
\setsymbol{LFI:FWHM:LFI25}{30.46}
\setsymbol{LFI:FWHM:LFI26}{30.31}
\setsymbol{LFI:FWHM:LFI27}{32.65}
\setsymbol{LFI:FWHM:LFI28}{32.66}



\setsymbol{LFI:FWHM:uncertainty:LFI18:units}{0.170\arcm}
\setsymbol{LFI:FWHM:uncertainty:LFI19:units}{0.174\arcm}
\setsymbol{LFI:FWHM:uncertainty:LFI20:units}{0.170\arcm}
\setsymbol{LFI:FWHM:uncertainty:LFI21:units}{0.156\arcm}
\setsymbol{LFI:FWHM:uncertainty:LFI22:units}{0.164\arcm}
\setsymbol{LFI:FWHM:uncertainty:LFI23:units}{0.171\arcm}
\setsymbol{LFI:FWHM:uncertainty:LFI24:units}{0.652\arcm}
\setsymbol{LFI:FWHM:uncertainty:LFI25:units}{1.075\arcm}
\setsymbol{LFI:FWHM:uncertainty:LFI26:units}{1.131\arcm}
\setsymbol{LFI:FWHM:uncertainty:LFI27:units}{1.266\arcm}
\setsymbol{LFI:FWHM:uncertainty:LFI28:units}{1.287\arcm}

\setsymbol{LFI:FWHM:uncertainty:LFI18}{0.170}
\setsymbol{LFI:FWHM:uncertainty:LFI19}{0.174}
\setsymbol{LFI:FWHM:uncertainty:LFI20}{0.170}
\setsymbol{LFI:FWHM:uncertainty:LFI21}{0.156}
\setsymbol{LFI:FWHM:uncertainty:LFI22}{0.164}
\setsymbol{LFI:FWHM:uncertainty:LFI23}{0.171}
\setsymbol{LFI:FWHM:uncertainty:LFI24}{0.652}
\setsymbol{LFI:FWHM:uncertainty:LFI25}{1.075}
\setsymbol{LFI:FWHM:uncertainty:LFI26}{1.131}
\setsymbol{LFI:FWHM:uncertainty:LFI27}{1.266}
\setsymbol{LFI:FWHM:uncertainty:LFI28}{1.287}


\setsymbol{HFI:center:frequency:100GHz:units}{100\,GHz}
\setsymbol{HFI:center:frequency:143GHz:units}{143\,GHz}
\setsymbol{HFI:center:frequency:217GHz:units}{217\,GHz}
\setsymbol{HFI:center:frequency:353GHz:units}{353\,GHz}
\setsymbol{HFI:center:frequency:545GHz:units}{545\,GHz}
\setsymbol{HFI:center:frequency:857GHz:units}{857\,GHz}

\setsymbol{HFI:center:frequency:100GHz}{100}
\setsymbol{HFI:center:frequency:143GHz}{143}
\setsymbol{HFI:center:frequency:217GHz}{217}
\setsymbol{HFI:center:frequency:353GHz}{353}
\setsymbol{HFI:center:frequency:545GHz}{545}
\setsymbol{HFI:center:frequency:857GHz}{857}


\setsymbol{HFI:Ndetectors:100GHz}{8}
\setsymbol{HFI:Ndetectors:143GHz}{11}
\setsymbol{HFI:Ndetectors:217GHz}{12}
\setsymbol{HFI:Ndetectors:353GHz}{12}
\setsymbol{HFI:Ndetectors:545GHz}{3}
\setsymbol{HFI:Ndetectors:857GHz}{4}


\setsymbol{HFI:FWHM:Maps:100GHz:units}{9\parcm88}
\setsymbol{HFI:FWHM:Maps:143GHz:units}{7\parcm18}
\setsymbol{HFI:FWHM:Maps:217GHz:units}{4\parcm87}
\setsymbol{HFI:FWHM:Maps:353GHz:units}{4\parcm65}
\setsymbol{HFI:FWHM:Maps:545GHz:units}{4\parcm72}
\setsymbol{HFI:FWHM:Maps:857GHz:units}{4\parcm39}
\setsymbol{HFI:FWHM:Maps:100GHz}{9.88}
\setsymbol{HFI:FWHM:Maps:143GHz}{7.18}
\setsymbol{HFI:FWHM:Maps:217GHz}{4.87}
\setsymbol{HFI:FWHM:Maps:353GHz}{4.65}
\setsymbol{HFI:FWHM:Maps:545GHz}{4.72}
\setsymbol{HFI:FWHM:Maps:857GHz}{4.39}


\setsymbol{HFI:beam:ellipticity:Maps:100GHz}{1.15}
\setsymbol{HFI:beam:ellipticity:Maps:143GHz}{1.01}
\setsymbol{HFI:beam:ellipticity:Maps:217GHz}{1.06}
\setsymbol{HFI:beam:ellipticity:Maps:353GHz}{1.05}
\setsymbol{HFI:beam:ellipticity:Maps:545GHz}{1.14}
\setsymbol{HFI:beam:ellipticity:Maps:857GHz}{1.19}


\setsymbol{HFI:FWHM:Mars:100GHz:units}{9\parcm37}
\setsymbol{HFI:FWHM:Mars:143GHz:units}{7\parcm04}
\setsymbol{HFI:FWHM:Mars:217GHz:units}{4\parcm68}
\setsymbol{HFI:FWHM:Mars:353GHz:units}{4\parcm43}
\setsymbol{HFI:FWHM:Mars:545GHz:units}{3\parcm80}
\setsymbol{HFI:FWHM:Mars:857GHz:units}{3\parcm67}

\setsymbol{HFI:FWHM:Mars:100GHz}{9.37}
\setsymbol{HFI:FWHM:Mars:143GHz}{7.04}
\setsymbol{HFI:FWHM:Mars:217GHz}{4.68}
\setsymbol{HFI:FWHM:Mars:353GHz}{4.43}
\setsymbol{HFI:FWHM:Mars:545GHz}{3.80}
\setsymbol{HFI:FWHM:Mars:857GHz}{3.67}


\setsymbol{HFI:beam:ellipticity:Mars:100GHz}{1.18}
\setsymbol{HFI:beam:ellipticity:Mars:143GHz}{1.03}
\setsymbol{HFI:beam:ellipticity:Mars:217GHz}{1.14}
\setsymbol{HFI:beam:ellipticity:Mars:353GHz}{1.09}
\setsymbol{HFI:beam:ellipticity:Mars:545GHz}{1.25}
\setsymbol{HFI:beam:ellipticity:Mars:857GHz}{1.03}


\setsymbol{HFI:CMB:relative:calibration:100GHz}{$\lsim 1\%$}
\setsymbol{HFI:CMB:relative:calibration:143GHz}{$\lsim 1\%$}
\setsymbol{HFI:CMB:relative:calibration:217GHz}{$\lsim 1\%$}
\setsymbol{HFI:CMB:relative:calibration:353GHz}{$\lsim 1\%$}
\setsymbol{HFI:CMB:relative:calibration:545GHz}{}
\setsymbol{HFI:CMB:relative:calibration:857GHz}{}


\setsymbol{HFI:CMB:absolute:calibration:100GHz}{$\lsim 2\%$}
\setsymbol{HFI:CMB:absolute:calibration:143GHz}{$\lsim 2\%$}
\setsymbol{HFI:CMB:absolute:calibration:217GHz}{$\lsim 2\%$}
\setsymbol{HFI:CMB:absolute:calibration:353GHz}{$\lsim 2\%$}
\setsymbol{HFI:CMB:absolute:calibration:545GHz}{}
\setsymbol{HFI:CMB:absolute:calibration:857GHz}{}


\setsymbol{HFI:FIRAS:gain:calibration:accuracy:statistical:100GHz}{}
\setsymbol{HFI:FIRAS:gain:calibration:accuracy:statistical:143GHz}{}
\setsymbol{HFI:FIRAS:gain:calibration:accuracy:statistical:217GHz}{}
\setsymbol{HFI:FIRAS:gain:calibration:accuracy:statistical:353GHz}{2.5\%}
\setsymbol{HFI:FIRAS:gain:calibration:accuracy:statistical:545GHz}{1\%}
\setsymbol{HFI:FIRAS:gain:calibration:accuracy:statistical:857GHz}{0.5\%}


\setsymbol{HFI:FIRAS:gain:calibration:accuracy:systematic:100GHz}{}
\setsymbol{HFI:FIRAS:gain:calibration:accuracy:systematic:143GHz}{}
\setsymbol{HFI:FIRAS:gain:calibration:accuracy:systematic:217GHz}{}
\setsymbol{HFI:FIRAS:gain:calibration:accuracy:systematic:353GHz}{}
\setsymbol{HFI:FIRAS:gain:calibration:accuracy:systematic:545GHz}{7\%}
\setsymbol{HFI:FIRAS:gain:calibration:accuracy:systematic:857GHz}{7\%}


\setsymbol{HFI:FIRAS:zero:point:accuracy:100GHz:units}{0.8\MJysr}
\setsymbol{HFI:FIRAS:zero:point:accuracy:143GHz:units}{}
\setsymbol{HFI:FIRAS:zero:point:accuracy:217GHz:units}{}
\setsymbol{HFI:FIRAS:zero:point:accuracy:353GHz:units}{1.4\MJysr}
\setsymbol{HFI:FIRAS:zero:point:accuracy:545GHz:units}{2.2\MJysr}
\setsymbol{HFI:FIRAS:zero:point:accuracy:857GHz:units}{1.7\MJysr}

\setsymbol{HFI:FIRAS:zero:point:accuracy:100GHz}{0.8}
\setsymbol{HFI:FIRAS:zero:point:accuracy:143GHz}{}
\setsymbol{HFI:FIRAS:zero:point:accuracy:217GHz}{}
\setsymbol{HFI:FIRAS:zero:point:accuracy:353GHz}{1.4}
\setsymbol{HFI:FIRAS:zero:point:accuracy:545GHz}{2.2}
\setsymbol{HFI:FIRAS:zero:point:accuracy:857GHz}{1.7}


\setsymbol{HFI:unit:conversion:100GHz:units}{0.2415\MJysrmK}
\setsymbol{HFI:unit:conversion:143GHz:units}{0.3694\MJysrmK}
\setsymbol{HFI:unit:conversion:217GHz:units}{0.4811\MJysrmK}
\setsymbol{HFI:unit:conversion:353GHz:units}{0.2883\MJysrmK}
\setsymbol{HFI:unit:conversion:545GHz:units}{0.05826\MJysrmK}
\setsymbol{HFI:unit:conversion:857GHz:units}{0.002238\MJysrmK}

\setsymbol{HFI:unit:conversion:100GHz}{0.2415}
\setsymbol{HFI:unit:conversion:143GHz}{0.3694}
\setsymbol{HFI:unit:conversion:217GHz}{0.4811}
\setsymbol{HFI:unit:conversion:353GHz}{0.2883}
\setsymbol{HFI:unit:conversion:545GHz}{0.05826}
\setsymbol{HFI:unit:conversion:857GHz}{0.002238}


\setsymbol{HFI:colour:correction:alpha=-2:V1.01:100GHz}{0.9893}
\setsymbol{HFI:colour:correction:alpha=-2:V1.01:143GHz}{0.9759}
\setsymbol{HFI:colour:correction:alpha=-2:V1.01:217GHz}{1.0007}
\setsymbol{HFI:colour:correction:alpha=-2:V1.01:353GHz}{1.0028}
\setsymbol{HFI:colour:correction:alpha=-2:V1.01:545GHz}{1.0019}
\setsymbol{HFI:colour:correction:alpha=-2:V1.01:857GHz}{0.9889}


\setsymbol{HFI:colour:correction:alpha=0:V1.01:100GHz}{1.0008}
\setsymbol{HFI:colour:correction:alpha=0:V1.01:143GHz}{1.0148}
\setsymbol{HFI:colour:correction:alpha=0:V1.01:217GHz}{0.9909}
\setsymbol{HFI:colour:correction:alpha=0:V1.01:353GHz}{0.9888}
\setsymbol{HFI:colour:correction:alpha=0:V1.01:545GHz}{0.9878}
\setsymbol{HFI:colour:correction:alpha=0:V1.01:857GHz}{1.0014}

\providecommand{\sorthelp}[1]{}

\title{\Spitzer\ \Planck\ \Herschel\ Infrared Cluster (SPHerIC) survey: Candidate galaxy clusters  at $1.3 < z < 3$ selected by high star-formation rate}
\authorrunning{C. Martinache, A. Rettura, H. Dole et al.}
\titlerunning{SPHerIC galaxy cluster candidates at $1.3 < z < 3$}

\author{C. Martinache\inst{1,2},
A. Rettura\inst{3}, 
H. Dole\inst{1} \and
M. Lehnert\inst{4} \and
B. Frye\inst{5} \and
B. Altieri\inst{6} \and
A. Beelen\inst{1} \and
M. Béthermin\inst{7} \and
E. Le Floc'h\inst{8,9} \and
M. Giard\inst{10} \and
G. Hurier\inst{11} \and
G. Lagache\inst{7} \and
L. Montier\inst{10} \and
A. Omont\inst{4} \and
E. Pointecouteau\inst{10} \and
M. Polletta\inst{7,10,12} \and
J.-L. Puget\inst{1} \and
D. Scott\inst{13} \and
G. Soucail\inst{10} \and
N. Welikala\inst{14}
}

\institute{Institut d'Astrophysique Spatiale, Univ. Paris-Sud, CNRS, Univ. Paris-Saclay, b\^at 121, Univ-Paris Sud, 91405 Orsay Cedex, France 
\and
Department of Astronomy, Universidad de Concepci\'on, Casilla 160-C, Concepci\'on, Chile
\and
IPAC, Caltech, KS 314-6, Pasadena, CA 91125, USA
\and
Sorbonne Universit\'{e}, CNRS UMR 7095, Institut d'Astrophysique de Paris, 98 bis bd Arago, 75014 Paris, France
\and 
Department of Astronomy/Steward  Observatory, 933 N.~Cherry Avenue, University  of  Arizona,  Tucson,  AZ  85721,
USA
\and
Herschel Science Centre, European Space Astronomy Centre, ESA, 28691 Villanueva de la Ca\~nada, Spain
\and
Aix-Marseille Univ., CNRS, Laboratoire d'Astrophysique de Marseille, 13013, Marseille, France
\and
IRFU, CEA, Universit\'e Paris-Saclay, F-91191 Gif-sur-Yvette, France
\and
Universit\'e Paris Diderot, AIM, Sorbonne Paris Cit\'e, CEA, CNRS, F-91191 Gif-sur-Yvette, France
\and 
IRAP (Institut de Recherche en Astrophysique et Plan\'etologie), Universit\'e de Toulouse, CNRS, UPS, Toulouse, France
\and
Centro de Estudios de Física del Cosmos de Aragón (CEFCA), Plaza de San Juan 1 Planta-2, 44001 Teruel, Spain
\and
INAF–Istituto di Astrofisica Spaziale e Fisica Cosmica Milano, via Bassini 15, 20133 Milano, Italy
\and
Department of Physics and Astronomy, University of British Columbia, 6224 Agricultural Road, Vancouver BC V6T 1Z1, Canada
\and
Sub-Department  of  Astrophysics,  University  of  Oxford,  Keble Road, Oxford OX1 3RH, UK}
        
\date{Submitted ... / Accepted ...}

\abstract{There is a lack of large samples of spectroscopically confirmed clusters and protoclusters at high redshifts, $z>$1.5.  Discovering and characterizing distant (proto-)clusters is important for yielding insights into the formation of large-scale structure and on the physical processes responsible for regulating star-formation in galaxies in dense environments.  The \Spitzer\ \Planck\ \Herschel\ Infrared Cluster (SPHerIC) survey was initiated to identify these characteristically faint and dust-reddened sources during the epoch of their early assembly.  We present \Spitzer\ /IRAC observations of 82 galaxy (proto-)cluster candidates at 1.3\,<\,$z_p$\,<\,3.0 that were vetted in a two step process: (1) using \Planck\ to select by color those sources with the highest star-formation rates, and (2) using \Herschel\ at higher resolution to separate out the individual red sources.  The addition of the \Spitzer\ data enables efficient detection of the central and massive brightest red cluster galaxies (BRCGs).  We find that BRCGs are associated with highly significant, extended and crowded regions of IRAC sources which are more overdense than the field.  This result corroborates our hypothesis that BRCGs within the \Planck\ - \Herschel\ sources trace some of the densest and actively star-forming proto-clusters in the early Universe. On the basis of a richness-mass proxy relation, we obtain an estimate of their mean masses which suggests our sample consists of some of the most massive clusters at z$\approx$2 and are the likely progenitors of the most massive clusters observed today.
}

               
\maketitle

\section{Introduction}\label{sect:intro}
 
Our understanding of the formation and evolution of galaxy clusters is linked to developing insights into the growth of large-scale structure and the way in which environment affects the growth of the ensemble of galaxies.
The fact that galaxies residing in the densest environments evolve differently compared to field galaxies
has been firmly established observationally at low and intermediate redshifts ($z \lesssim$ 1), with tight
correlations between galaxy density and various galaxy properties 
\citep[see e.g.,][]{Dressler1980, Gomez2003, Balogh2004, Kauffmann2004, Peng2010, Cooper2010, Muzzin2012}. 

Among other relationships between galaxy properties and environment, the star-formation rate (SFR)--density
relation shows that, at least out to $z$$\approx$1, galaxies in the field are more actively forming stars,
whereas galaxies in dense environments have finished forming the bulk of their stars \citep{Gomez2003}.
This tendency is reversed at higher redshifts \citep{Elbaz2007, Rodighiero2007}. Many systems at
$z$$\gsim$1.5 do indeed show indications of vigorous episodes of star-formation in the cores of the densest
environments \citep{Tran2010, Hilton2010, Hayashi2010, Santos2015, Wang2016}. However, somewhat surprisingly,
evolved systems are also found at these redshifts \citep[see e.g.,][]{Gobat2011, Andreon2014}, and some
studies report no inversion of the relation, making the nature of this reversal unclear
\citep[e.g.,][]{Quadri2012}.  

Similarly, the tight color-magnitude relation shows the predominance of red massive galaxies with little
ongoing star formation in dense environments. This implies that massive cluster galaxies show little change
in their properties over the redshift range 0 to $\approx$1.4 and are simply evolving passively
\citep{Bower1992, Kodama1997, Hogg2003, Mei2009}. The situation should be drastically different at
$z$$\ga$1.5-2. Indeed, most studies are consistent with a model in which massive cluster galaxies formed the
bulk of their stellar population at high-$z$ \citep[$z_f$$\sim$3-5;][]{Bower1992, Eisenhardt2008}, and have
since evolved passively.  Moreover, although in the present-day universe cluster galaxies contribute less than 1\% to the
cosmic SFR density, it is expected theoretically that they contribute $\approx20\%$ at
redshifts $z$$\sim$2--3, and up to $50\%$ at $z$$\sim$10 \citep{Bethermin2013, Bethermin2014, Popesso2015a, Popesso2015b, Chiang2017, Muldrew2018}.
The study of clusters and  of objects that will become massive and virialized by $z$$\approx$0, or protoclusters, is thus vital for understanding the global 
star-formation history of the Universe.

Many questions remain about the physical processes governing and affecting the growth of cluster galaxies,
and therefore direct in situ observations of clusters in the early Universe are needed to understand
galaxy evolution. Unfortunately, currently, only $\sim$100 clusters have been found at $z$$>$1
\citep{Muzzin2013}, and a few handfuls are spectroscopically confirmed at $z$$>$1.5 \citep{Overzier2016, noirot16, noirot18}. The relatively small size of the
sample of $z>1.5$ clusters precludes us from obtaining a better understanding of the role of environment in
galaxy evolution on the growth of galaxies.   It
is therefore not surprising that significant effort over the past years has been expended in constructing
reliable, well-controlled samples of galaxy clusters at the higher redshifts. 

Since clusters are rare objects on the sky, we rely on large surveys to identify overdensities 
of galaxies or particular types of galaxies expected to lie predominately in clusters (e.g., powerful
high redshift radio galaxies).
Several selection techniques are generally used, each having different limitations and biases. Overall,
several methods have had been particularly successful in discovering clusters and overdensities.
For example, overdensities can be detected as hot ionized plasmas either from extended X-Ray emission
\citep{Tozzi2015, Rosati2002, Rosati2004, Fassbender2011, Vikhlinin2006} or Sunyaev-Zel'dovich (SZ) effect
\citep{Planck_XXIV, Brodwin2010, Bleem2015, Marriage2011}. Overdense regions have also been discovered through
galaxy counts in optical or IR bands or in making color slices in multiband optical and IR surveys
\citep{Papovich2008, Rettura2014, Gobat2011, Brodwin2013, Andreon2014, Muzzin2013, Papovich2010}. 
We note that at $z\gtrsim 2$, we do not necessarily expect protoclusters to be virialized and hence they 
are not expected to have significant X-Ray emission or produce a measureable SZ decrement. 

Moreover, because of limitations
in the various instruments used for these surveys, most studies are limited to  $z$$\la$1.5.
X-ray surveys are limited in redshift by the fact that
X-ray luminosity quickly decreases with redshift ($\propto\!(1+z)^4$) and are not currently 
of large enough aperture to identify such faint clusters. The SZ signals are, in principle,
independent of redshift, but the resolution of current instruments limits the detection to the most massive
systems, and thus effectively restricts the detection of the SZ signal to $z$$\la$1 
for \Planck\ \citep{Planck_XXIV}, and up to $z$$\la$1.4 for the SPT 
\citep[see e.g.,][]{Bleem2015, Stanford2012}. 

Large area near- and mid-infrared surveys have also been successful in constructing samples of clusters or candidates clusters at $z\!>\!1$. Such studies take advantage of the fact that the cosmological dimming which increases with redshift is partially compensated by the relatively steep slope of the spectral energy distribution (SED) in the near and mid-infrared which results in a  ``negative $k-$correction''. The $K-$correction is such that the observed flux density at 4.5 microns is almost constant over the redshift range $z$$\sim$0.7--2.5 \citep{Wylezalek2013}. The approach to find protocluster candidates is to search for  old, passive stellar populations in wide angle near and mid-IR surveys \citep{Papovich2008, Rettura2014}. This is done by using colors to identify the emission peak at $1.6~\mu$m which appears in populations dominated by stars older than $\sim$10 Myrs.  This technique can then be used to select galaxies over a particular broad redshift range that depends on the filters used, and is, for mid-infrared filters, typically 1$<z<$3.
Although candidate clusters selected in this way require the additional step of spectroscopic confirmation,  this method is
both well-constrained and systematic. 

These studies from the literature provide a solid methodology that we can build and expand upon to construct a sample of robust high-$z$ 
(proto)clusters candidates from the original \Planck\ selection \citep{Planck_XXXIX} and subsequent follow-up with \Herschel\ \citep{Planck_XXVII}. The approach used in these two \Planck\ studies was to select cluster-sized highly star-forming regions using \Planck\ photometry.  The specific combinations used in the photometry selects extragalactic sources within the range of $z$ = 1-4. 
These follow-up \Herschel\ studies also provide strong evidence that the vast majority of objects in the catalog of bright, color-selected \Planck\ sources are nascent clusters at $z>1.5$, which are generally not selected efficiently or in as great a number in other surveys. As a result, these high-$z$ clusters have
different galaxy densities, sizes, morphologies, and SEDs.
So this approach of selecting the largest scale, most extremely star-forming regions over the whole sky, which is largely independent of any other property of the galaxies in overdensities, is likely an efficient way of discovering bona-fide protoclusters.  

\begin{figure}[!ht]
\begin{center}
\includegraphics[width=0.45\textwidth]{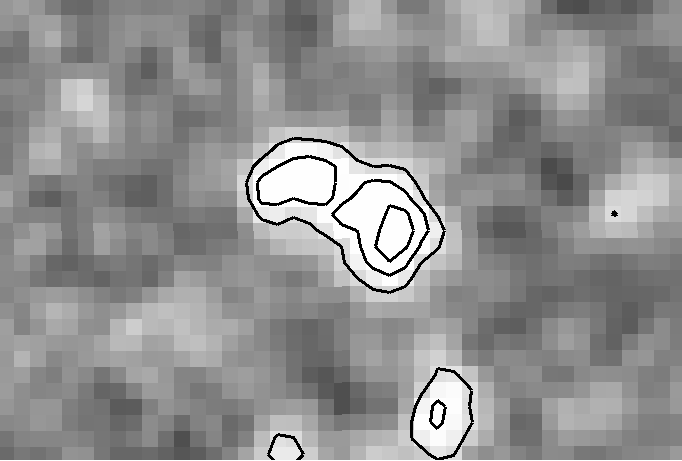}
\includegraphics[width=0.45\textwidth]{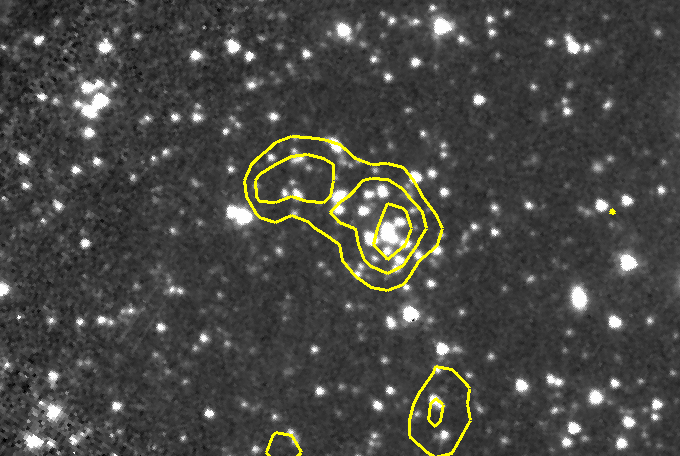}
\caption{\emph{Top}: A $4.2\arcmin \times 3\arcmin$ \Herschel /SPIRE $350~\mu$m image of one SPHerIC survey field. Black contours indicate increasing emission levels at $3,4,5 \sigma$ about the RMS noise of the image. \emph{Bottom}: The \Spitzer/IRAC $3.6~\mu$m image of the same field. Yellow contours indicate the same SPIRE $350~\mu$m emission levels as above. The gain in angular resolution in using IRAC at $3.6~\mu$m is a 
factor $\sim15$ better than SPIRE at $350~\mu$m.
Thus, the IRAC imaging permits
the identification of many individual galaxies within the SPIRE beam, as shown.}
\label{fig:compResol}
\end{center}
\end{figure}

To further elucidate if these objects are in fact (proto)clusters and determine what are the
photometric properties of their constituent galaxies, we obtained warm \Spitzer\ observations at 3.6 and 4.5 microns with the mid-infrared camera, IRAC \citep{Fazio2004, Werner2004}. In particular, we 
followed-up 89 out of the 216 fields that were
acquired previously with \Herschel/SPIRE. In this study, we present the analysis of the IRAC data of these 89 sources and their interpretation. The two
bands observed contain information about the stellar content of objects within each SPIRE field, and the IRAC
angular resolution allows us to separate out the individual SPIRE sources
(Fig.~\ref{fig:compResol}). Color-selection using the IRAC bands further allows us to 
select putative (proto)cluster members and investigate their colors, estimate their photometric redshifts, and calculate their level of spatial clustering \citep[e.g.,][]{Papovich2008, Rettura2014}.

This paper is organized as follows: in Section~\ref{sect:spheric} we describes the \Planck/\Herschel\ selection of our candidate clusters; Section~\ref{sect:data} presents the observations, data reduction, catalog construction and completeness of our survey; Section \ref{sect:cluster_det} presents the analysis of the density and colors of the sources in each field. Finally, we discuss the composition of the \Spitzer\ \Planck\ \Herschel\ Infrared Clusters (SPHerIC) sample in Section~\ref{sect:Discussion} and  summarize our conclusions in Section~\ref{sect:conclusions}. 

Throughout this paper we have used magnitudes in the AB system unless indicated otherwise.  
We assume a flat $\Lambda$CDM model, with cosmological parameters from the \Planck\ 2015 release \citep{Planck_XIII}: $\Omega_{\Lambda} = 0.691$ ; $\Omega_{m} = 0.308 $ ; $\Omega_{b} = 0.0486 $ ; $H_0 = 67.8 $ km s$^{-1}$ Mpc$^{-1}$. We adopt a Salpeter initial mass function. 

\section{SPHerIC survey}\label{sect:spheric}

In this paper, we present a sample of high-$z$ (proto)cluster candidates observed with the \Spitzer\ space telescope. This is a subsample of a parent sample that was built from observations with the \Planck\ \citep{planck2014-a01} and \Herschel\ \citep{Pilbratt2010}. The parent sample was discussed in great detail in \cite{Planck_XXXIX} and \cite{Planck_XXVII}. We call this subsample the \textit{\textbf{S}pitzer} \textit{\textbf{P}lanck \textbf{Her}schel} \textbf{I}nfrared \textbf{C}lusters (SPHerIC) survey sample.  

\subsection{\Planck/HFI selection: The PHz sample}\label{subsect:phz}

To discover galaxy (proto)clusters in their phase of most intense stellar mass growth, the approach adopted in \cite{Planck_XXXIX} was to use \Planck\ photometry to select candidates using various color combinations. This approach was chosen specifically to select highly star-forming extragalactic sources on Mpc-scales in the redshift range, $z$=1 to 4. The \Planck\ satellite, particularly its High Frequency Instrument (HFI), was uniquely capable of performing such a search because of three major advantages.


First, the observing frequency range of HFI of 100~857 GHz and is distributed across this range in 6 photometric bands\footnote{with central frequencies of 100, 143, 217, 353, 545 \& 857 GHz and with relative bandpasses in each band of $\frac{\Delta f }{f}\!\sim\!30\%$.},
is well-suited for detecting the redshifted emission peak of the thermal dust continuum over the redshift range, $z$=1 - 4. Dust emission is a direct tracer of star formation, and the many possible color combinations of
the \Planck\ photometry are  effective in separating thermal emission from distant galaxies from other astrophysical contaminants such as cosmic infrared background, cosmic microwave background anisotropies, galactic cirrus, cold galactic sources, synchrotron emission, and low-$z$ galaxies.


Secondly, the beam size of HFI naturally selects Mpc-scale regions at these redshifts; for example at 545 GHz the \Planck\ PSF FWHM is $4.5\arcmin$, corresponding to 2.3~Mpc at $z=2$. 

Finally, the all-sky coverage of \Planck\ means that we can perform a systematic search for rare and
extremely bright regions of infrared emission. We note, however, that the work presented in \cite{Planck_XXXIX} only considered the cleanest 26\% of the extragalactic sky.


A sample of 2151 \Planck\ high-$z$ candidates is constructed using this selection technique. We designate this sample as the \textit{\textbf{P}lanck} list of \textbf{H}igh-\textbf{\textit{z} }sources \citep[PHz;][]{Planck_XXXIX}. The PHz, as constructed, is expected to contain $z$$\ga$1.5 (proto)clusters in the major mass assembly stage of their most massive galaxies, and also includes a small minority of lensed galaxies and chance line-of-sight alignments of submm sources \citep[see, e.g.,][]{Canameras2015, Flores-Cacho2016, Negrello2017}. 

Estimates of the redshifts and star-formation rates of the PHz objects using the far-Infrared (FIR) photometry
result in a redshift distribution that peaks at $z=2.5$, with 90\% of the sources expected to be in the
redshift range $1.5-3.7$, and in SFRs that are $\sim$10$^4~M_{\odot} \mathrm{yr}^{-1}$. See Section 6 of
\cite{Planck_XXXIX} for a full description.  Although these estimates indicate that many of the \Planck\
sources could be distant (proto)clusters, low spatial resolution and possible confusion of the \Planck\
photometry implies that these data along are insufficient to constrain the exact nature of these sources. The
subsequent higher resolution \Herschel\ follow-up observations at least partially overcomes these limitations.

\subsection{\Herschel/SPIRE observations}\label{subsect:hpasss}

The Herschel Space Observatory ( \Herschel\ ) provides a unique opportunity to further investigate the candidates from 
the PHz. Indeed, the wavelength of the bands in SPIRE,  $250, 350, 500\mu$m \citep{Griffin2010}, overlap with
two of the bands of \Planck/HFI, 545 and 857 GHz.  Fortunately, the far better sensitivity and angular
resolution of \Herschel\, which is a factor of eight times better at 545 GHz compared to \Planck/HFI, makes it possible to identify the
counterparts contributing to the far-infrared emission of the PHz sources.  
Most of the 232 fields observed with SPIRE show unambiguous overdensities of galaxies \citep{Planck_XXVII},
and a few, $\sim 5$\% of the sample, show exceptionally bright lensed systems 
\citep{Planck_XXVII, Canameras2015}. Given that \Herschel\ observations have a $98\%$ success rate in detecting galaxy sources, with only $2\%$ of the 
detections due to Galactic cirrus emission, our novel \Planck\ selection is demonstrated to be efficient.  In total, 216 candidate 
galaxy overdensities were discovered with \Herschel\ . These SPIRE sources are typically exceptionally
bright, red, and highly clustered, suggesting that these are in fact good candidates for 
(proto)clusters in their most intense phase of star formation. 
Such objects have been detected in large \Herschel\ surveys  
\citep[e.g.,][]{Clements2014, Clements2016, Oteo2018}, but we stress that the \Planck\ selection results
in the detection of meaningful signposts to pinpoint good protocluster candidates
\citep[see also][]{Greenslade2018}.  

Despite finding large density contrasts and red colors of the associated SPIRE sources, the SPIRE data alone do not conclusively determine the nature of the \Planck\ sources.  This is in part because we cannot exclude the possibility of chance alignments of bright infrared galaxies. We have therefore undertaken a large multiwavelength photometric and spectroscopic follow-up campaign to constrain the nature of our sources. Here we have focused on  follow-up with the Spitzer Space Telescope ( \Spitzer\ ). 

\subsection{\Spitzer/IRAC target selection}
\label{subsect:spheric}

\Spitzer/IRAC \citep{Fazio2004} introduces two major advantages to the further investigation of the candidates of the PHz:

(i) \Spitzer/IRAC covers the rest-frame red optical to near-infrared at z$\ga$1, a key wavelength range to probe the stellar content of galaxies over the expected redshift range of the \Planck/\Herschel\ selected sources. The characteristics of the spectral energy distributions in the rest-frame near-infrared are such that the two bands are well-placed for selecting potential protocluster members over the expected redshift range.

(ii) \Spitzer/IRAC enables the study our protocluster candidates at arcsecond resolution, making it possible to resolve the individual galaxy members. In contrast, the $\sim18\arcsec$ beam of SPIRE smears out the signals of individual galaxy members such that only the light integrated up over to several sources is detected. 
In Fig.~\ref{fig:compResol} we select one example to demonstrate the gain in resolution in IRAC images compared to those from SPIRE. 

We requested warm \Spitzer/IRAC observations at 3.6 and 4.5 microns for some of the most promising PHz sources which had also been also observed by \Herschel/SPIRE (see Sect.~\ref{subsect:spitzer_obs} for details).  At the time the first two proposals were submitted, a director's discretionary time (GO) and a guest observer proposal in cycle 9 (GO9), the \Herschel\ program was only partially completed. Thus the target fields were selected based on the  availability of SPIRE data. Unfortunately, the subsample observed as part of these programs randomly samples the distribution of density contrast of red SPIRE sources, $\delta_\textrm{SPIRE, red}$ and the significance of their overdensities, $\sigma_{\delta, \mathrm{SPIRE, red}}$. For the guest observer call in cycle 11 (GO11), targets were chosen so that the density contrast of red SPIRE sources (See Sect. \ref{subsect:hpasss}) was $\delta_{\rm SPIRE, red}>3$ \textbf{or} the significance of the over-density was $\sigma_{\delta, \mathrm{SPIRE, red}}>15$ \citep{Planck_XXVII}. We thus expect that our IRAC sub-sample is representative of the parent sample, although perhaps slightly biased toward the most overdense sources.

In total, 89 fields were observed with \Spitzer. Out of these 89 fields, we discarded five fields as the
images have a bright star within the field of view preventing detection and accurate photometry; and we also removed two fields because they are not covered in both IRAC photometric channels available during the warm \Spitzer\ mission. The sample presented here thus contains 82 candidate galaxy overdensities. We call this subsample, that has \Spitzer\ observations, the \textbf{S}pitzer \textit{\textbf{P}lanck \textbf{Her}schel} \textbf{I}nfrared \textbf{C}lusters (SPHerIC) sample.  

\begin{figure}[!ht]
\centering
\includegraphics[width=0.45\textwidth]{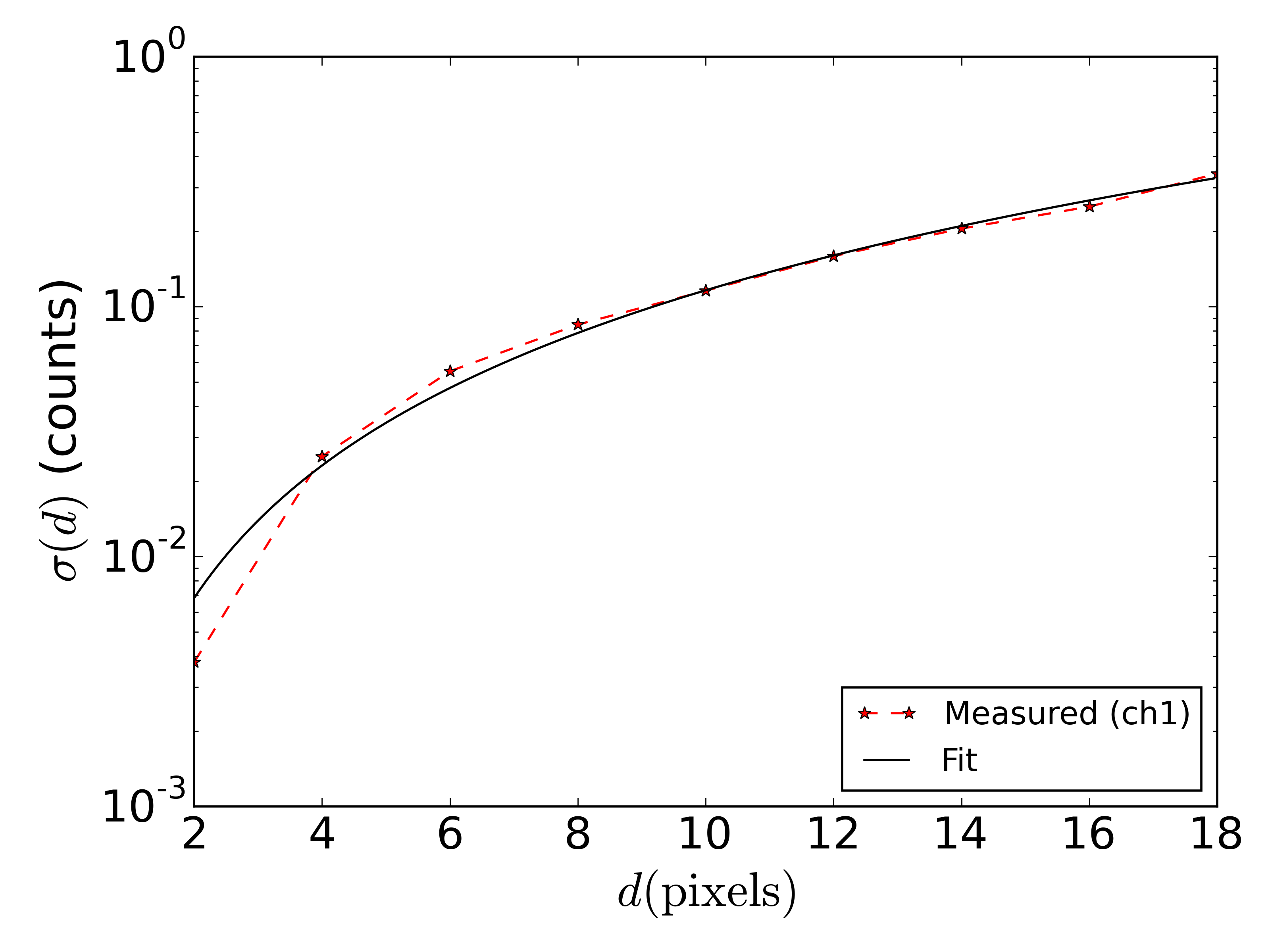}
\caption{RMS measured in apertures, $\sigma(d)$, in the 3.5$\mu$m \Spitzer\ images as a function of aperture diameter, $d$, in units of pixels. The solid line is a fit to the data with Eq.~\ref{eq:rms_r_PL}, as is done in \cite{Gawiser2006}.}
\label{fig:rms_vs_d}
\end{figure}

\section{Data}
\label{sect:data}

\subsection{Observations}
\label{subsect:spitzer_obs}

\Spitzer/IRAC \citep{Fazio2004} observations at 3.6 and $4.5~\mu$m were performed during the warm mission in three programs: PID 11004 (4 sources in a DDT, PI: H. Dole); 90111 (33 sources in GO9, PI: H. Dole) and 80238 (52 sources in GO11, PI: H. Dole). Our observing strategy was to take 100 second integrations per exposure, with a 12-point medium cycle dither over all exposures, resulting in a net integration time of about 1200s per sky pixel. We used post-basic calibrated data (PBCD) mosaics provided by the \Spitzer\ Science Center (SSC) in our analysis. For the complete analysis, we only consider the area of the final images that have at least 500s per pixel of exposure yielding a usable area for each field of $\approx$5$\arcmin$$\times$5$\arcmin$ at $3.6~\mu$m (hereafter sometimes refereed to as channel 1, or ch1) and $4.5~\mu$m (hereafter sometimes refereed to as channel 2, or ch2).  There are also two additional side fields that are covered only in one band. In this anaysis, we have not considered these additional side-fields. The common sky area covered by both channels is well matched to the angular size of the regions of interest, since it covers one \Planck\ beam of FWHM of 4.5\arcmin\ \citep{planck2014-a01,planck2014-a09}, and a few \Herschel/SPIRE sources that are often within beam of the \Planck\ source \citep{Planck_XXVII}.

\subsection{Source extraction and photometry}
\label{subsect:photometry}

The source extraction is performed using SExtractor, in dual image mode using the 4.5 microns mosaic as the
detection image \citep{Bertin1996}. We used the parameters from \cite{Lacy2005} which are optimized for the
detection of faint sources in IRAC images and use the coverage maps as weight maps. The \textbf{initial} detection threshold is set to $2 \sigma$. The same parameters are used for both ch1 and ch2. 
Flux densities are measured in $4\arcsec$ diameter apertures, optimizing the signal-to-noise ratio (SNR) of detected sources. Mosaics are converted from MJy sr$^{-1}$to $\mu$Jy pix$^{-1}$ using a conversion factor of 8.4615 $\mu$Jy pix$^{-1}$ (MJy sr$^{-1}$)$^{-1}$, for our pixel scale of 0$\arcsec$.60012. 
Aperture correction is then applied to the returned flux densities from SExtractor. Aperture correction factors are derived using IRAC PSF images constructed from observations of stars in the EGS field \citep{Barmby2008}, using the flux density in a $24\arcsec$ diameter aperture as a reference (Appendix~\ref{app:aper_cor}). In this way, we obtained correction factors of 1.37 for ch1 and 1.39 for ch2, consistent with values presented in \cite{Barmby2008}, \cite{Ashby2009}, and \cite{Wylezalek2013}.

We evaluated the completeness of our data by inserting artificial point sources in the real images. In each of the final mosaics, 50 sources of the same flux density are inserted, as to keep the change in surface density below $5\%$. This procedure is repeated 15 times in each mosaic, and for 30 logarithmically-spaced flux density values from $0.1\mu$Jy to $200\mu$Jy. Once the completeness is estimated, the IRAC catalogs are then truncated to only include sources that are above the $50\%$ completeness limit in the $4.5\mu$m channel (2.54$\mu$Jy or 22.9 AB mag). Determining the completeness is useful for testing whether or not our analysis is sensitive to the exact significance cut used in selecting the sources.  To do this, we performed the same analysis with an 80\% completeness cut and the main results remained unchanged.

\subsection{Correction of SExtractor photometric error}
\label{subsect:sex_err_corr}

\begin{figure}[!ht]
\begin{center}
\includegraphics[width=0.45\textwidth]{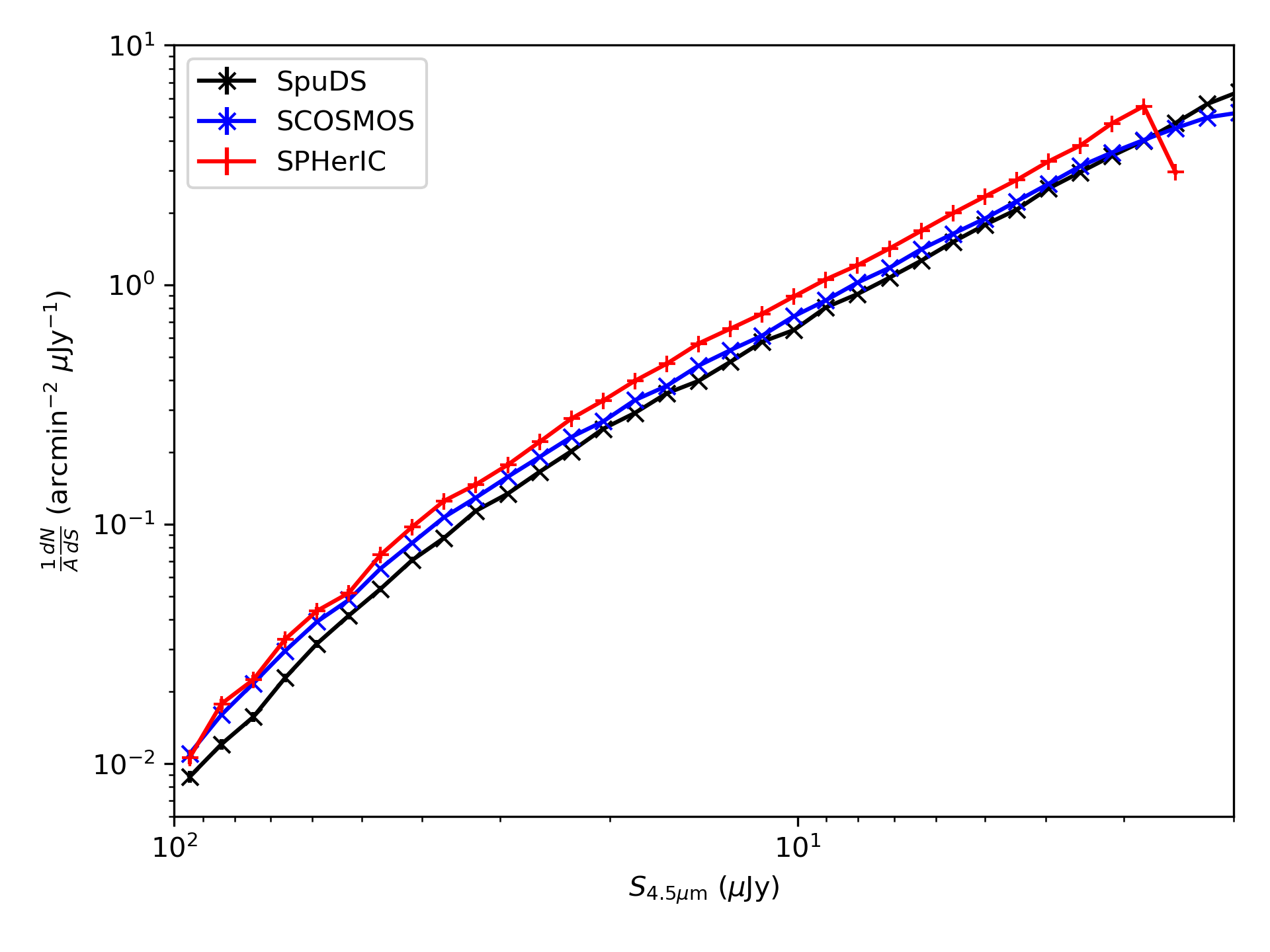}
\caption{Differential number counts for sources detected at $4.5\mu$m in our SPHerIC survey (red line) compared to the SpUDS (black line) and SCOSMOS (blue line) results. Error bars are computed using Poissonian statistics and do not take cosmic variance into account.}
\label{fig:nb_counts}
\end{center}
\end{figure}

SExtractor assumes a Poisson uncorrelated pixel noise to derive photometric uncertainties. However, this is not the case for our resampled, averaged PBCD mosaics. Because of the difficulty of knowing and modeling all the sources of systematic and random errors \citep[beyond just detector and confusion noise, see e.g.,][]{Dole2003}, we characterized the noise empirically \cite[specifically following the method presented in][]{Gawiser2006}. In this empirical method we measured the flux in apertures of radius $r$, with $r$ ranging from $1.5\arcsec$ to $15\arcsec$. The apertures were placed at random in regions free of sources. Regions that are devoid of sources were selected using the SExtractor segmentation map. Sixty apertures were placed in each candidate field. The total flux distribution was then computed and the different values of the background were accounted for in each field. A Gaussian function was then fitted to this distribution giving an estimate of the RMS in each map for a range of aperture sizes in images in both IRAC bands. A power law was then fitted to the obtained relation for the noise as a function of the radius of the aperture, $\sigma(r)$,

\begin{equation}
\label{eq:rms_r_PL}
\sigma(r) = \sigma_1 \alpha r^{\beta}
\end{equation}

\noindent
where $\sigma_1$ if the pixel-pixel RMS noise (i.e., aperture of 1 pixel), and $\alpha$ and $\beta$ are the fitted coefficients of the power-law. The fitted curve is shown in Fig.~\ref{fig:rms_vs_d} and the fitted values of the coefficients are provided in Table~\ref{tab:fitted_coeffs_rms}.

\begin{table}
\caption{\label{tab:fitted_coeffs_rms} Fitted parameters of $\sigma(r)$ (Eq.~\ref{eq:rms_r_PL}). See also Fig.~\ref{fig:rms_vs_d}.} 
\centering
\begin{tabular}{lcc}
 \hline 
 \hline 
 coefficient & ch1 & ch2 \\ 
 \hline 
 $\sigma_1$ (counts) & 0.00697 & 0.0068 \\ 
 $\alpha$ & 0.98 & 1.32 \\ 
 $\beta$ & 1.76 & 1.56 \\ 
 \hline 
 \end{tabular}  
\end{table}

\noindent
The final SExtractor photometric uncertainties are corrected for correlated noise as a function of aperture size, $\sigma_{corr} (r)$, as,

\begin{equation}
\sigma_{corr} (r) = \left(  \frac{\sigma^2(r)+ \frac{F}{G}}{\sigma_1^2 \pi r^2 + \frac{F}{G}} \right)^{\frac{1}{2}} \sigma_{SE}(r)
,\end{equation}

\noindent
where $F$ is the source flux (in counts), $G$ is the instrumental gain\footnote{G = $\mathtt{GAIN} \times \mathtt{EXPTIME} \times \mathtt{TOTALBCD}/\mathtt{FLUXCONV}$; these values are provided in the file headers of the PBCD mosaics.} and $\sigma_{SE}$ is the uncertainty given by SExtractor \citep[see][for details]{Gawiser2006}. This estimate accounts for the correlation between the pixels when deriving photometric uncertainties. 

\begin{figure}[ht]
\includegraphics[width=0.48\textwidth]{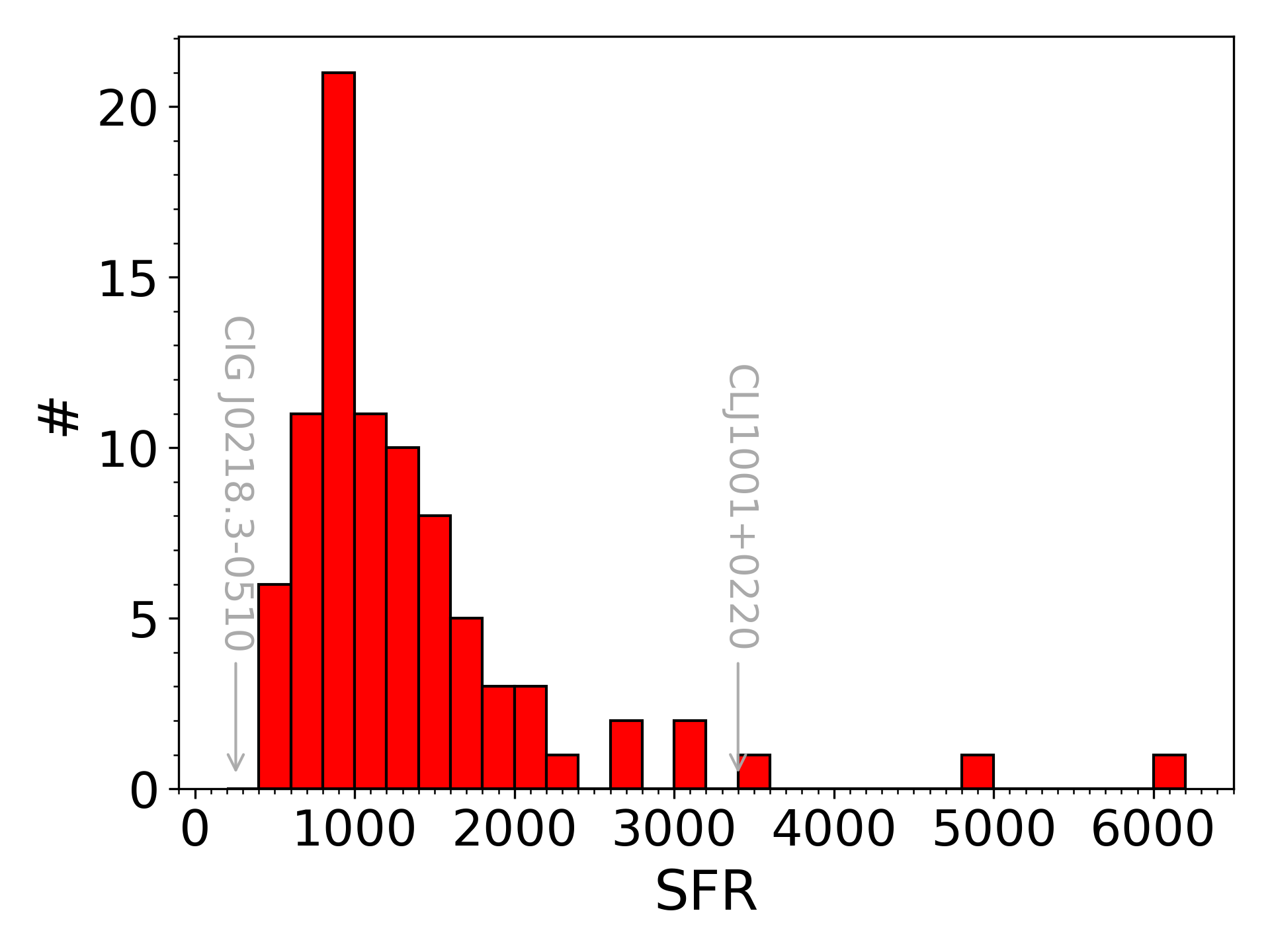}
\caption{Distribution of the SFRs of the 82 BRCGs of our sample. SFRs were estimated from infrared luminosities $L_\mathrm{IR}$ using the relation of \cite{Bell2003}. We estimated $L_\mathrm{IR}$ by fitting a modified blackbody with a fixed dust emissivity spectral index, $\beta$=1.5, and a fixed dust temperature, $T_\mathrm{dust}=35$~K, to the \Herschel\ photometric points. The only other free parameter was the redshift of the sources $z$ \citep[see][for details]{PlanckXXVIII2014}.
For comparison with our estimated SFRs, we show the total SFR in the 80\,kpc core of CL J1001+0220 at $z=2.506$ \citep{Wang2016} and the estimated SFR of the brightest \Herschel\ source in the core of ClG J0218.3-0510, a protocluster located at $z=1.62$ \citep{Santos2014}. It is interesting to note that the total SFR in the core of CL J1001+0220 is comprised of
nine luminous star-forming galaxies that are resolved from a single bright \Herschel\ SPIRE source \citep{Wang2016}.}
\label{fig:sfr_spire_brcg}
\end{figure}

\subsection{Number counts}
\label{subsect:nb_counts}

In order to test and validate the accuracy of our photometry, we calculate galaxy differential number counts. We use the SpUDS and SCOSMOS \Spitzer\ data as control samples. SpUDS (P.I.: J.Dunlop) and SCOSMOS \citep{Sanders2007} both cover an area of $\sim1$ deg$^2$, and have deeper flux density limit than the SPHerIC survey. 

To treat the different samples in a homogeneous way, we used SExtractor in single mode to extract sources in the SPHerIC, SpUDS, and SCOSMOS fields.  The parameters are set to be the same parameters used in \cite{Lacy2005}.   We did not use public catalogs since those are truncated to a flux density limit that is too high to make a fair comparison (8$\sigma$ for SpUDS, 5$\sigma$ for SCOSMOS). We note that our extractions are consistent with the results in the publicly available catalogs. We see an excess of sources in the SPHerIC survey, especially at low flux densities. This is to be expected if our hypothesis that our fields contain overdensities of galaxies is valid (see Fig.~\ref{fig:nb_counts}). However, we caution that we did not consider the impact of cosmic variance when computing the uncertainties and thus we are underestimating the uncertainties in the number counts for each of the fields in all flux bins.
 
\section{\textit{Planck/Herschel} highly star-forming sources as seen by \Spitzer}
\label{sect:cluster_det}

\begin{figure*}[!ht]
\includegraphics[width=1.0\textwidth]{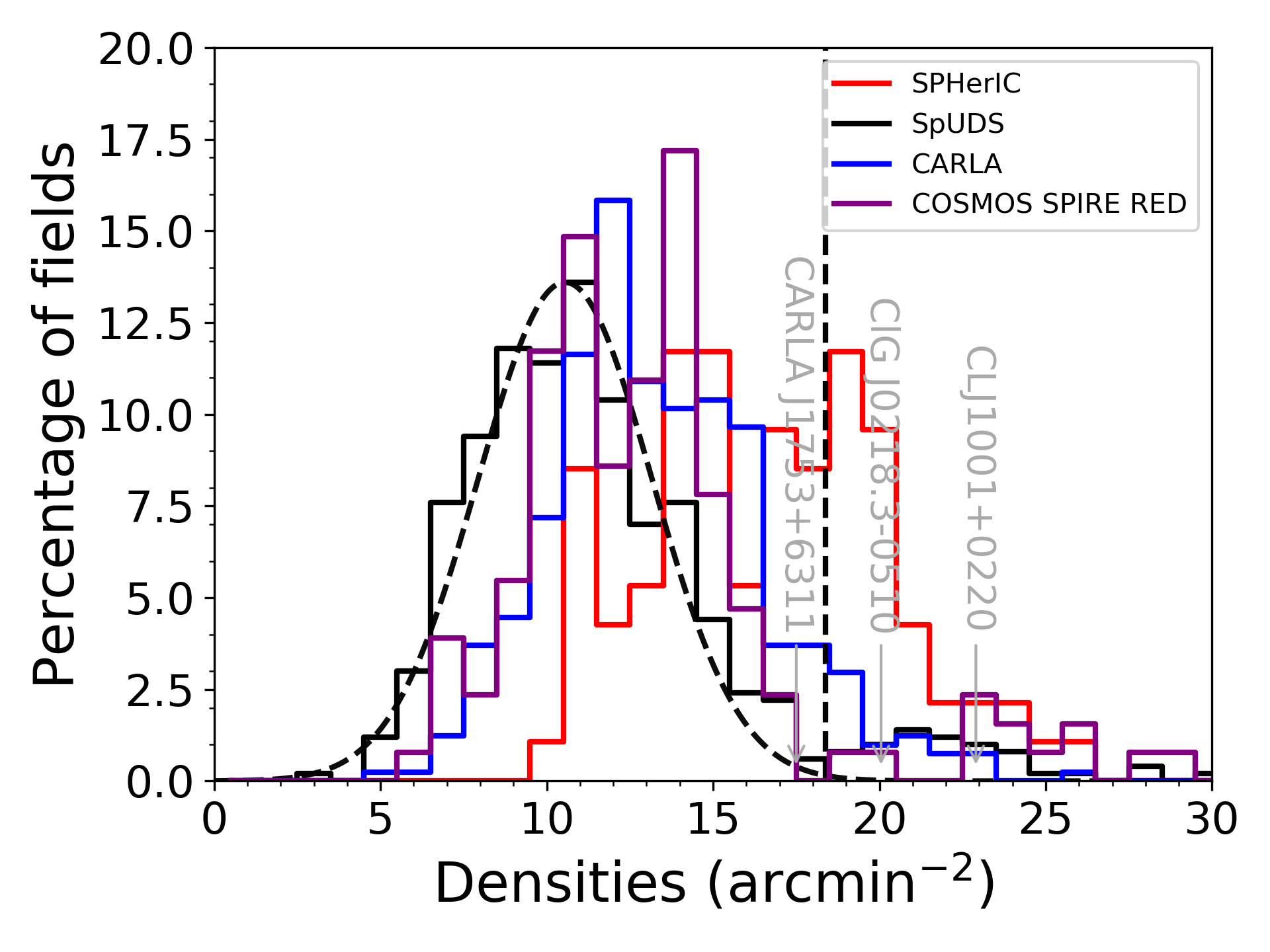}
\caption{Distribution of surface densities of IRAC red sources in the SPHerIC sample (red histogram). These surface densities were estimated within a 1\arcmin\ radius around each BRCG. For comparison, we show the surface density of random 1\arcmin\ radius fields within the SpUDS images (black histogram). We fitted a Gaussian distribution to the lower half of the SpUDS distribution (black dashed curve). The blue histogram shows the distribution of surface densities from the CARLA cluster survey \citep{Wylezalek2013}. To compare our sample with random red SPIRE sources, selected in the same way as for the BRCGs in the SPHerIC sample, we estimated the surface density of red IRAC sources within 1\arcmin\ radius of red SPIRE sources in the COSMOS field (purple histogram). 
The vertical black dashed line indicates the densities that are 3$\sigma$ above the mean of the SPUDS control field, a reasonable line for demarcating overdense sources on scales of 1 arc min. 
We note that 46\% of the SPHerIC candidates are above this threshold compared to only 9\% of the CARLA fields. As a reference, we also indicate in gray the surface densities of three other \Spitzer\ selected spectroscopically confirmed clusters: CARLA J1753+6311, a cluster at $z=1.58$ \citep[][Rettura et al. 2018, in prep.]{Cooke2016}, ClG J0218.3-0510, a protocluster at $z=1.62$ \citep{Papovich2010}, and CLJ1001+0220 at $z=2.506$ \citep{Wang2016}.}
\label{fig:density_distribution}
\end{figure*}

\begin{figure}[!ht]
\includegraphics[width=0.5\textwidth]{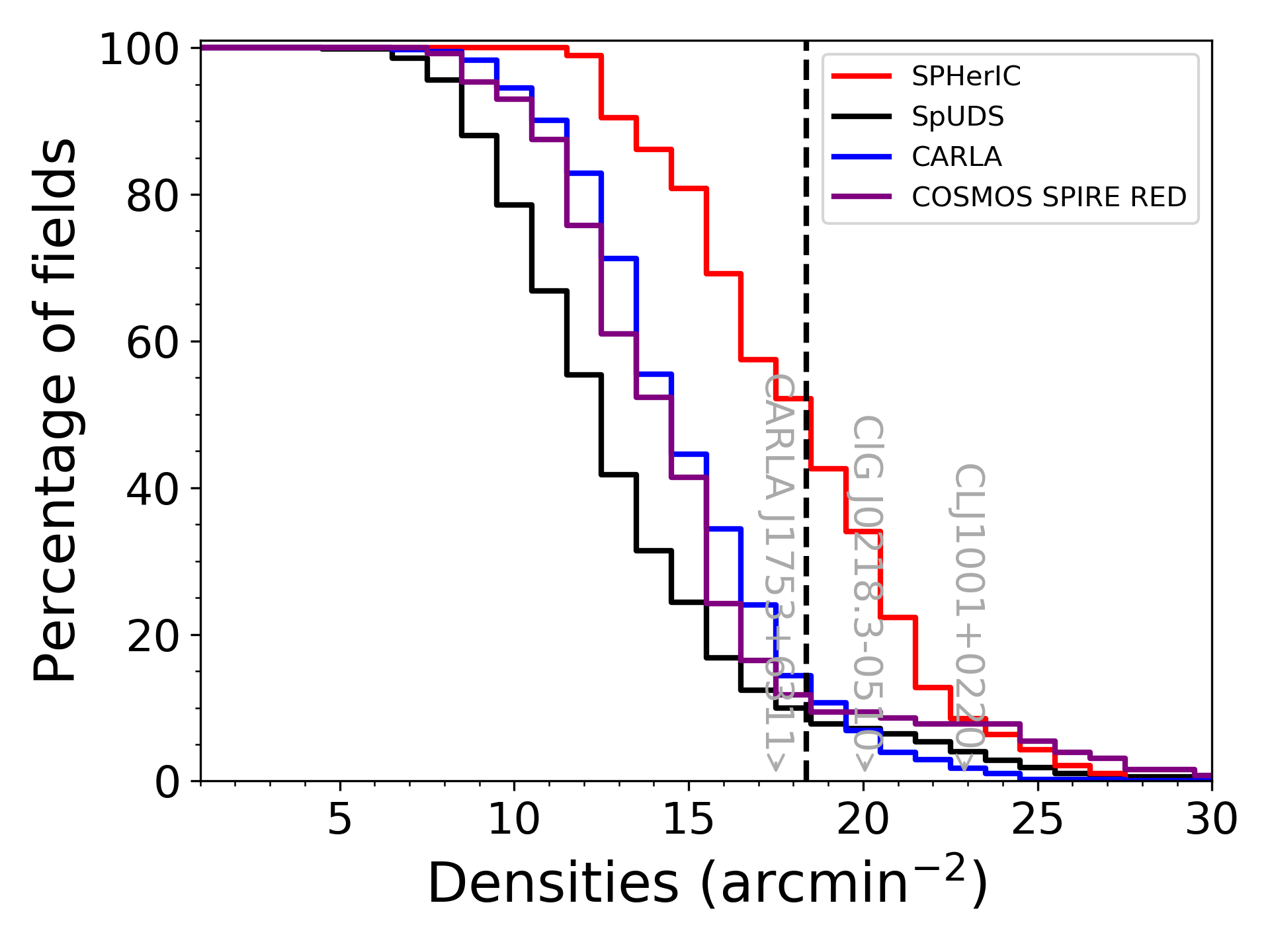}
\caption{Cumulative surface density distributions for the SPHerIC, SpUDS and CARLA surveys. Lines are the same as in Fig.~\ref{fig:density_distribution} and are indicated in the legend.}
\label{fig:density_distribution_cumul}
\end{figure}

\subsection{Brightest red cluster galaxy}
\label{subsect:BRCG}

\begin{figure*}[!ht]
\begin{center}
\includegraphics[width=0.8\textwidth]{{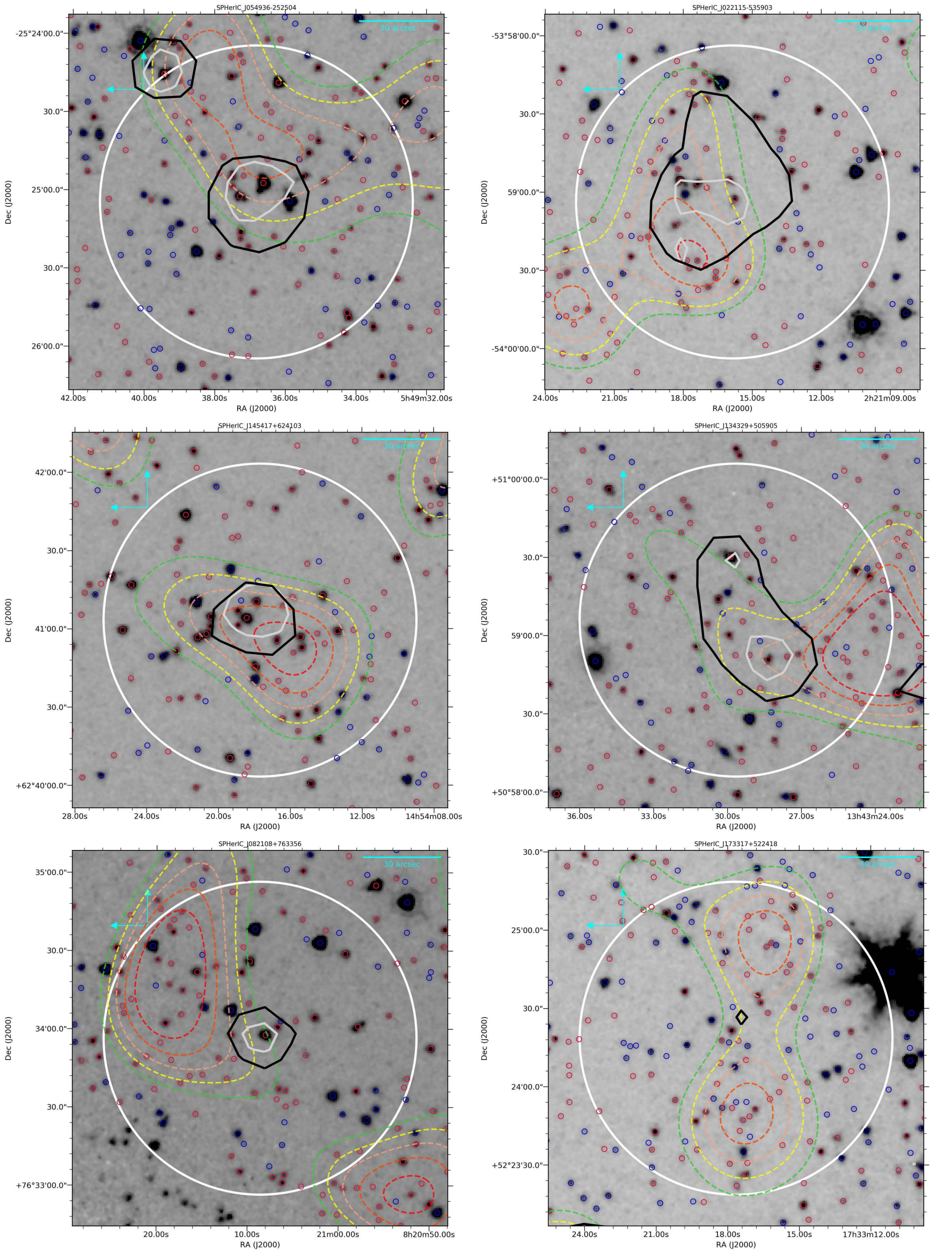}}
\caption{Images of the SPHerIC candidates at 4.5$\mu$m which are 4$\sigma$ overdense within a radius of 1\arcmin. Each image is 
$\sim\!2.4\arcmin\!\times\!2.4\arcmin$. IRAC sources with $[3.6]$--$[4.5]$ $>(<) -0.1$ are indicated by red (blue) circles. Contours of significance of the overdensities are shown green-yellow-salmon-orange-red lines starting at $2.5\sigma$ and incremented in $0.5\sigma$ steps. Herschel emission $3\sigma$ isocontours at 350 and 250$\mu$m are shown as black and gray lines respectively. The large white circle has a radius of $1\arcmin$ and is centered on the peak of emission at $350\mu$m from SPIRE.}
\label{fig:top12_1}
\end{center}
\end{figure*}

\begin{figure*}[!ht]
\begin{center}
\includegraphics[width=0.8\textwidth]{{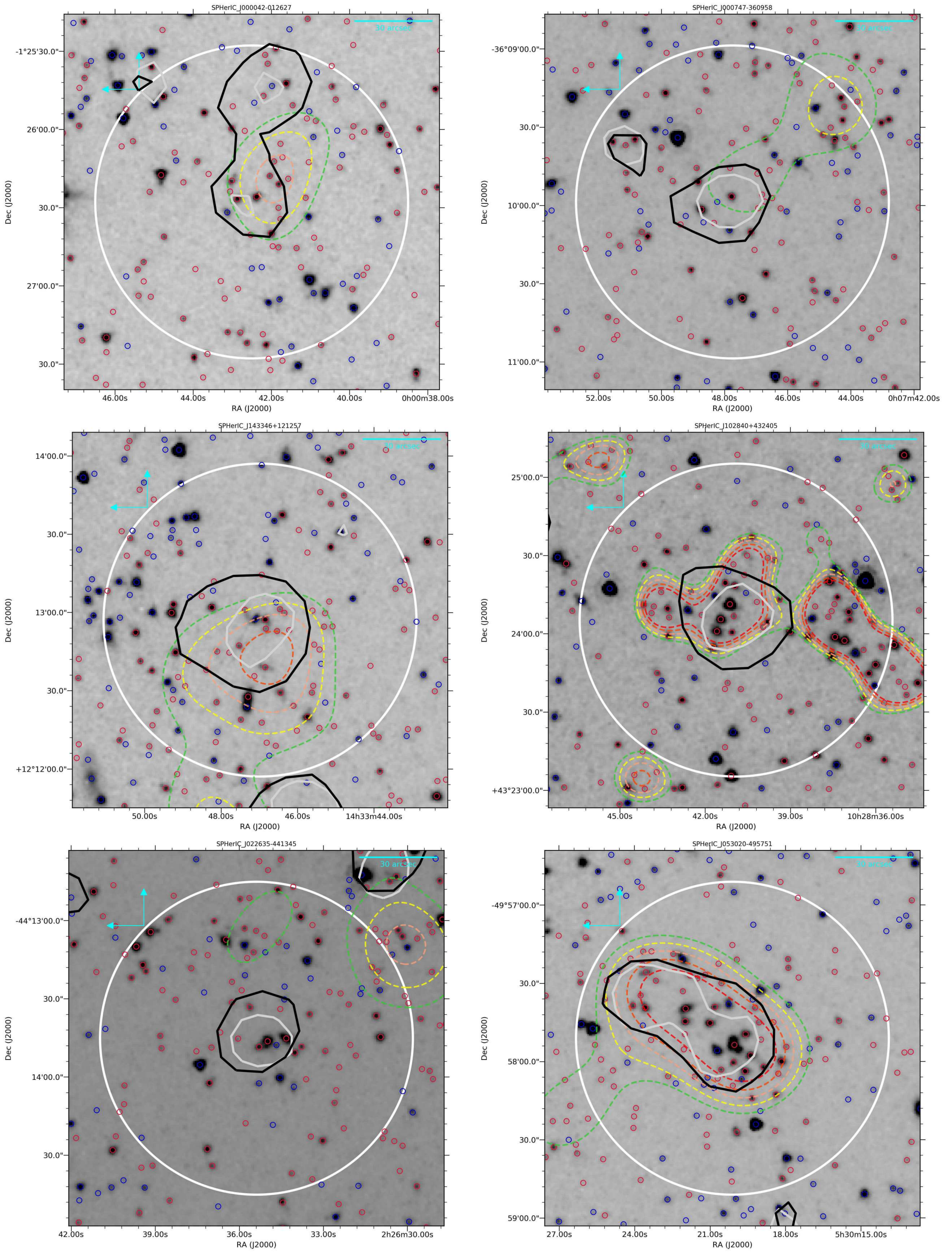}}
\caption{Continued from Fig.~\ref{fig:top12_1}.}
\label{fig:top12_2}
\end{center}
\end{figure*}


Each IRAC pointing covers approximately the \Planck\ source size and contains several \Herschel/SPIRE sources. 
Our main goal is to determine if the extremely high star-formation rates estimated in \Planck\ and 
\Herschel\ data are associated with galaxy overdensities and, if possible, measure the contamination 
rate of chance
alignments of several luminous star-forming galaxies at different redshifts. As we show below,  it is a challenge with the available data to accurately define the center of any putative overdensity.
Therefore, as a first assumption, we assumed that the brightest red SPIRE source (defined below) in each IRAC field represents the ``center'' of any potential overdensity. See Sect. \ref{subsect:AKDE} for a discussion of this assumption.

In each field, we estimated the average density of red IRAC sources (described in the next section) within of $1\arcmin$ of the brightest $350\mu$m red SPIRE source in the field. A red SPIRE source is defined as having relative flux densities in 3 SPIRE bands such that $S_{350}/S_{250} >0.7$ and $S_{500}/S_{350}>0.6$, which preferentially selects luminous IR emitters at z$\ga$2 \citep{Planck_XXVII}. These criteria for color selecting SPIRE sources is suitable for selecting heavily obscured galaxies over the expected redshift range of the \Planck\ sources in the parent catalog.
The bright, color-selected SPIRE source is therefore the brightest red cluster galaxy (hereafter, BRCG) in that field. The SPIRE photometry of these BRCGs indicate that they have very high star-formation rates (Fig.~\ref{fig:sfr_spire_brcg}). We note that the $350\mu$m flux densities of the BRCGs are typically less than the $350\mu$m flux densities of the confirmed Planck lenses \citep{Canameras2015}, and we estimate that at most 5-10\% of the remaining sample could be made up of lenses objects based on their flux densities relative to known lenses from various other surveys \citep[see e.g.,][and references therein]{Negrello2010, Negrello2017, Vieira2010, Wardlow2013, Nayyeri2016}. The angular size of $1\arcmin$ was chosen on physical grounds as it corresponds to a physical size of $\sim 500$kpc at $z=2$, roughly equal to $R_{500}$ for galaxy clusters at these redshifts \citep{Fassbender2011}. On more practical grounds, such a diameter around the BRCG fits  within the size of the \Spitzer\ images, and has the advantage of being easily reproducible in other surveys.

We note that in making the assumption that the BRCG represents the center of any possible overdensity, we are also testing whether or not infrared emission detected by \Planck\ and \Herschel\ yields a meaningful way to select galaxy (proto)clusters. This approach is conservative in that there may be overdense regions in the \Planck\ beam that are not associated directly with the brightest SPIRE source (see Sect.\ref{subsect:AKDE}). However, we note that this choice does have the advantage of being easily reproducible and allows us to rather straightforwardly compare our results with those of other surveys. 
Similar to the approach taken in the study of \cite{Wylezalek2013}, who targeted known active galactic nuclei at high redshifts to find  (proto)clusters, we aim to investigate whether BRCGs associated with \Planck\ sources are signposts of the presence of overdensities of high-$z$ (defined here as $z>1.3$) galaxies. Thus, our choice makes it straightforward to compare our results with those of other surveys that use particular classes of objects as indicators of overdensities.

\subsection{Selection of cluster member galaxies: IRAC red sources}\label{subsect:irac_red_sources}

In order to identify high redshift cluster member galaxies we applied a single IRAC color-cut: $[3.6]-[4.5]>-0.1$ \citep{Papovich2008}.  We will refer to these sources as IRAC red sources. Indeed, all realistic models of stellar populations older than 10 Myrs have a peak in the rest-frame spectral energy distribution at 1.6~$\mu$m with a power-law decline to longer wavelengths. This feature is widely used to estimate NIR photometric redshifts \citep[see e.g.,][]{Sawicki2002, Sorba2010}. Due to its relative insensitivity to the age of stellar populations, this feature is present almost completely independently of the evolutionary stage and type of the galaxy. \cite{Papovich2008} has shown this IRAC color-selection is able to efficiently select $z>1.3$ galaxies, with an $80\%$ success rate. The main contaminants are star-forming galaxies at $z\sim$0.3, powerful AGNs at all redshifts, and cool brown dwarfs. 
We note that we do not expect the contamination from these last type of objects to be significant in our study as we are selecting objects far away from the galactic plane. The \Planck\ selection is biased against the powerful Radio-Loud AGNs (a veto at 100~GHz was applied), but faint AGNs or radio-quiet AGNs could be present in our sample. 

The use of this color-criterion has already proven to be effective in constructing large samples of high-$z$ galaxy clusters, by searching for overdensities of such sources on the sky \citep{Galametz2012, Muzzin2013, Wylezalek2013, Rettura2014}. For example, \cite{Rettura2014} has shown that this color selection efficiently selects previously identified clusters at $z>1.3$ in the Bo$\ddot{\mathrm{o}}$tes survey discovered using an independent selection technique. Moreover, \cite{Papovich2010} discovered a $z=1.62$ structure with this technique, which was subsequently spectroscopically confirmed and is one of the best studied densest environments at $z>1.5$.

\subsection{Surface density estimates}
\label{subsect:surface_dens_est}

To estimate the degree of clustering and the galaxy density contrast around the BRCGs, we measure the surface densities of red IRAC sources around their positions, and compare the resulting distribution to reference samples. In order to derive surface density estimates, we performed a counts-in-cell analysis \citep{Wylezalek2013, Rettura2014}. We counted the number of IRAC red sources in circles of radius $1\arcmin$ centered on the BRCGs (see Sect. \ref{subsect:BRCG}). The resulting surface differential and cumulative density distributions of our sample are shown in Figs.~\ref{fig:density_distribution} and ~\ref{fig:density_distribution_cumul}. We compared our distribution to estimates from blank fields derived from the SpUDS photometry cut at the same limiting magnitudes as in our fields (m$_{AB}$=22.9 in the 4.5$\mu$m IRAC band). To derive
the estimates of the field density, we placed $500$ non-overlapping circles in the SpUDS mosaics and estimated the surface density of IRAC red sources in each of these regions
(Figs.~\ref{fig:density_distribution} and \ref{fig:density_distribution_cumul}). We find that the distributions are asymmetric skewing toward high galaxy surface densities. This is to be expected as the SpUDS field obviously contains some galaxy overdensities. Notably, SpUDS contains a $z=1.62$ confirmed protocluster \citep{Papovich2010}. To remove this tail in the galaxy density distribution, we fit a Gaussian distribution on the lower surface density half of the distribution \citep[e.g.,][]{Papovich2008, Galametz2012}. From this fit, we obtained an average galaxy surface density, $<\Sigma_{\mathrm{SpUDS}}> = 10.6$ arcmin$^{-2}$ and standard deviation, $\sigma_{\mathrm{SpUDS}} = 2.6$ arcmin$^{-2}$, comparable to the values found by \cite{Wylezalek2014}. Our the surface density distribution around the BRCGs of the SPHerIC sample is broader, $\sigma_{\mathrm{SPHerIC}} = 3.7 $ arcmin$^{-2}$, and peaks at higher surface densities, $<\Sigma_{\mathrm{SPHerIC}}> = 17.7$ arcmin$^{-2}$, compared to the surface density distribution of the SpUDS fields. We find that all our (proto)cluster candidate fields show a surface density higher than the mean surface density, and that 46\% (15\%, 7\%) of the SPHerIC fields are 3$\sigma$ (4, 5 $\sigma$) overdense compared to the SpUDS mean density and its standard deviation. 

Compared to similar observations obtained as part of the CARLA survey, \cite{Wylezalek2014} obtain a surface density distribution peaking at lower densities ($\Sigma_{\mathrm{CARLA}}\!\sim\!13.8 $ arcmin$^{-2}$ than the SPHerIC sample (cf. their Figure 2; Figs.~\ref{fig:density_distribution} and \ref{fig:density_distribution_cumul}). We note that 9\% (3\%) of the CARLA fields are 3 (4) $\sigma$ overdense compared to the mean density of the SpUDS fields. Their choice of targeting radio-sources to discover overdensities of galaxies is of course different from our use of BRCGs, but it is still noteworthy that our candidates exhibit on average even higher surface densities than the CARLA sample.
In Fig \ref{fig:top12_1}, we show the 12 candidates from the SPHerIC sample that have a density of red IRAC sources which are 4$\sigma$ greater than the mean surface density of the SpUDS field. The properties of these 12 candidates are enumerated in Table~\ref{tab:12_best}.  

Furthermore, to test if the overdensity is a consequence of estimating the density around a red SPIRE source and not related to our \Planck\ selection, we selected SPIRE sources in the COSMOS field that met our color criteria.  We performed exactly the same color cut to select red IRAC sources as we did for our SPHerIC sample and estimated the density of red IRAC sources within 1\arcmin of the red SPIRE sources in the COSMOS field.  We find that the surface and cumulative density distributions of the red IRAC sources around red SPIRE sources in the COSMOS field are similar to those found in the CARLA fields (Figs.~\ref{fig:density_distribution} and \ref{fig:density_distribution_cumul}). This suggests that while looking for overdensities generally around red SPIRE sources is perhaps a successful strategy, these overdensities are typically less that what we found for the SPHerIC sample \citep[see][]{Bethermin2014}. 

\begin{table*}
\centering
\caption{Twelve SPHerIC survey protocluster candidates with the highest surface densities of IRAC red sources surrounding a BRCG. All of these candidates are 4$\sigma$ above the mean surface density derived from SpUDS and are rank ordered by surface density (not RA). $a$ : discussed in \cite{Flores-Cacho2016}.}
\label{tab:12_best}
\begin{tabular}{lccc}
\hline
\hline
Name & RA & Dec & $\Sigma_{1'}$ \\
       & hh:mm:ss.ss & dd:mm:ss.s & arcmin$^{-2}$ \\
\hline
SPHerIC J053020$-$495751     & 05:30:20.11 & $-$49:57:51.1 & 26.42 \\
SPHerIC J102840$+$432405     & 10:28:40.91 & $+$43:24:05.2   & 25.78 \\
SPHerIC J000747$-$360958     & 00:07:47.75 & $-$36:09:58.7 & 24.83 \\
SPHerIC J022635$-$441345     & 02:26:35.39 & $-$44:13:45.1 & 24.51 \\
SPHerIC J143346$+$121257     & 14:33:46.98 & $+$12:12:57.1   & 23.87 \\
SPHerIC J000042$-$012627$^a$ & 00:00:42.50 & $-$01:26:27.8 & 23.55 \\
SPHerIC J173317$+$522418     & 17:33:17.65 & $+$52:24:18.6   & 22.60 \\
SPHerIC J134329$+$505905     & 13:43:29.65 & $+$50:59:06.0   & 22.28 \\
SPHerIC J022115$-$535903     & 02:21:15.86 & $-$53:59:03.8 & 21.96 \\
SPHerIC J082108$+$763356     & 08:21:08.60 & $+$76:33:56.5   & 21.96 \\
SPHerIC J145417$+$624103     & 14:54:17.68 & $+$62:41:03.3   & 21.33 \\
SPHerIC J054936$-$252504     & 05:49:36.81 & $-$25:25:04.8 & 21.01 \\
\hline\end{tabular}

\end{table*}

\subsection{Surface density profiles}\label{subsect:dens_profiles}

\begin{figure}[!ht]
\includegraphics[width=0.48\textwidth]{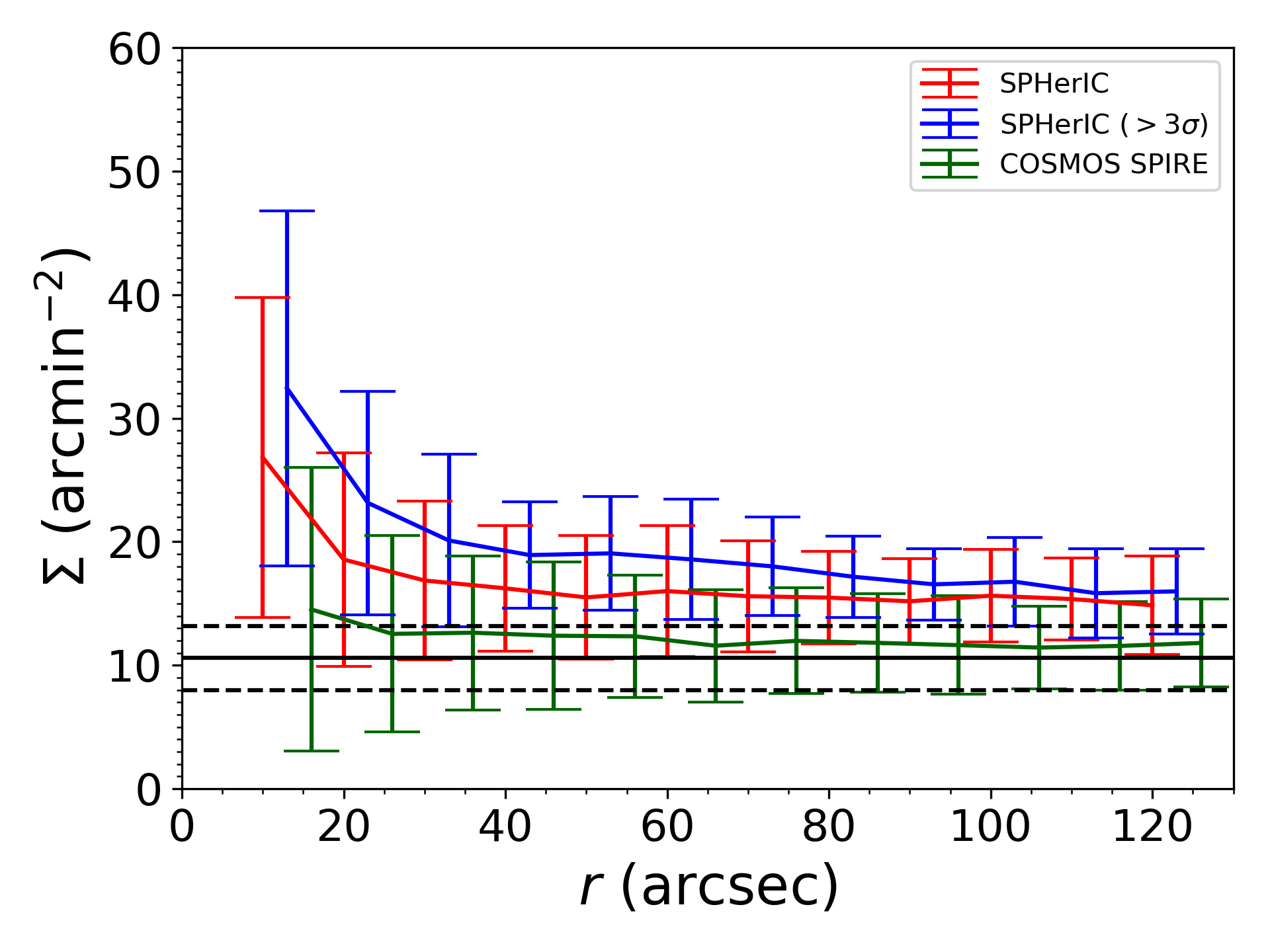}
\caption{Surface densities of red IRAC sources estimated in annuli of width $10\arcsec$ centered on the BRCGs.  The stack for the entire SPHerIC sample (82 candidates) is indicated by the red line, while the blue line is for the densest candidates (with $\Sigma_{60''} > \Sigma_{SpUDS}+ 3\sigma$). In green we show the profile around red SPIRE sources in the COSMOS field. The solid black line indicates the mean surface density $\Sigma_{\rm SpUDS}$ of IRAC red sources in the SpUDS survey, derived from a Gaussian fit as described in the text, while the dashed black lines indicate $\Sigma_\textrm{SpUDS}\pm \sigma_\textrm{SpUDS}$. Error bars indicate the scatter and not the error on the mean value. The blue and green lines are  offset in angular separation for clarity. 
}
\label{fig:density_profile}
\end{figure}

In this section, we investigate how the surface density of IRAC red sources depends on the distance
from the BRCG. In each field, we measured the surface density of IRAC red sources $\rho(r_1<r<r_2)$
in annuli of width $10\arcsec$ centered on the BRCG, with a minimum radius, $r_{min}=5.\arcsec$ and
a maximum radius, $r_{max}=125\arcsec$. We then stacked  the resulting profiles and conduct a similar analysis for
random positions in the SpUDs survey and around red SPIRE sources in the COSMOS field (Fig.~\ref{fig:density_profile}). 

The SpUDS density profile shows no radial dependence, consistent with an average region containing no significant overdensity (Fig.~\ref{fig:density_profile}). On the contrary, the SPHerIC fields show a strong rise in surface density at $r \leq 30\arcsec$. This behavior is enhanced when considering only the densest SPHerIC fields, that is, those with  $\Sigma > \Sigma_{\mathrm{SpUDS}}+3\sigma$. This implies that the density at scales $r\leq 30\arcsec$ and $1\arcmin$ are highly correlated, and that the average IRAC field centered on the BRCG is significantly overdense.  This demonstrates that the BRCG is indeed a good indicator of overdensities of red IRAC sources and lies approximately at the center of overdensities on scales of $\sim$0.5\arcmin.   

The radial profile around red SPIRE sources in the COSMOS field shows a mean density profile at all radii that is comparable to the density profile derived from the SpUDS fields although with a weak trend of some sources showing overdensities at small radii \citep[$\la$ 30 arcseconds; see][]{Bethermin2014}. This shows that red SPIRE sources in blind surveys do not generally have overdensities of IRAC selected galaxies, again corroborating our hypothesis that the \Planck\ detection and the BRCGs are particularly good beacons for discovering overdensities. 

It is also important to note is that the surface density is higher at all radii compared to the reference SpUDS sample. This is consistent with the fact that high-$z$ (proto)clusters are spatially extended, with angular sizes $\gtrsim2\arcmin$ \citep{Venemans2007, Chiang2013, Muldrew2015}. Given that our \Spitzer\ images do not physically extend beyond this radius, we expect our fields to have higher galaxy surface densities over the entire image. This is consistent with what has been
found surrounding high-$z$ radio-loud AGNs \citep[][their Fig.~6]{Wylezalek2013}. A dedicated program to cover a wider area  around each SPHerIC field is needed to study the radial surface density profiles of our sample in detail. 

\begin{figure}[!ht]
\includegraphics[width=0.48\textwidth]{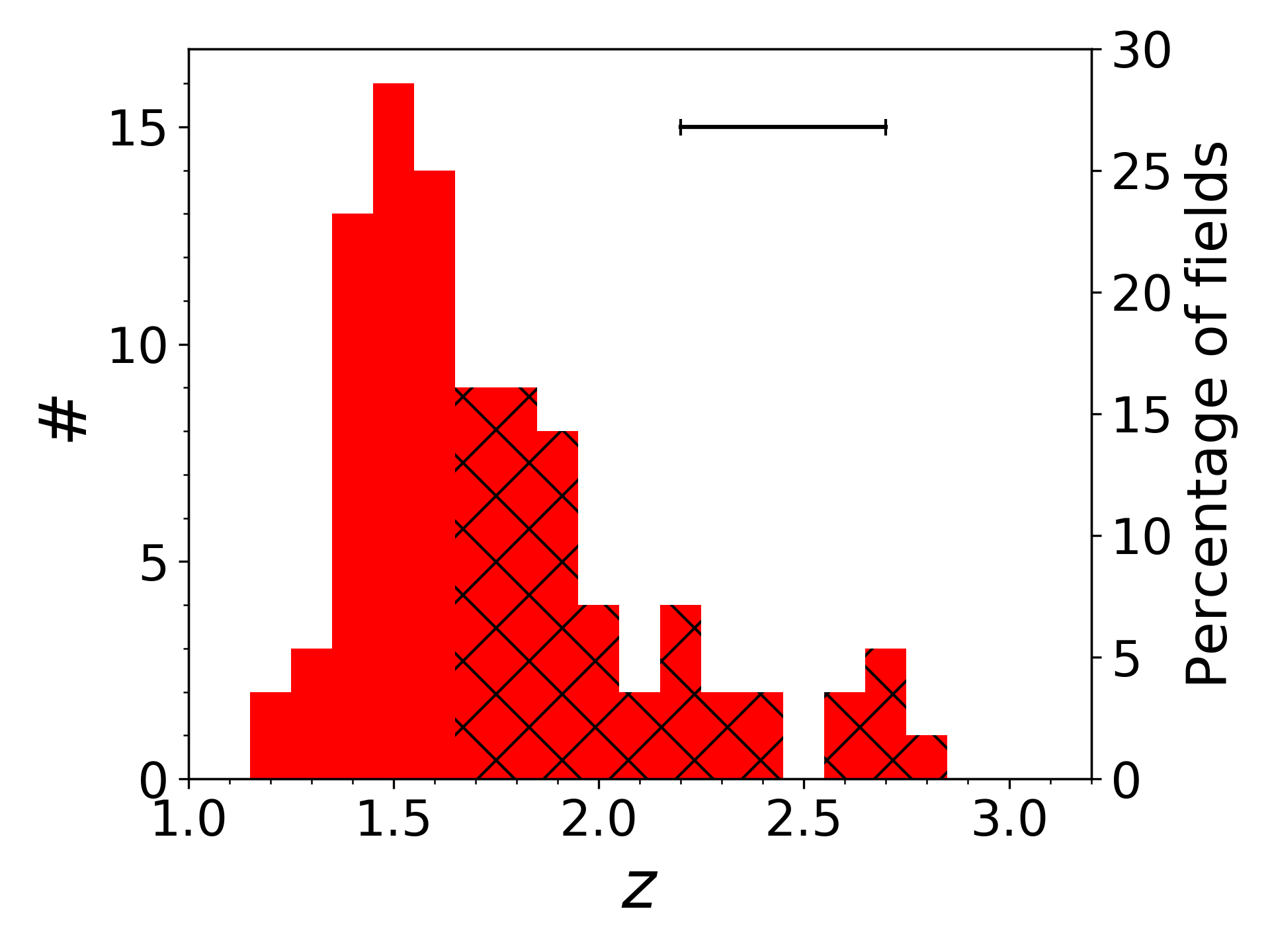}
\caption{Photometric redshift distribution derived from two band IRAC photometry of the SPHerIC cluster candidates. The hatched bars indicate the redshift above which the relationship used to derive the photometric redshift starts to break down.}
\label{fig:redshift_distribution}
\end{figure}

\section{Discussion}\label{sect:Discussion}

\subsection{Photometric redshift distribution}\label{subsect:phot-z}

We photometrically constrain the redshift distribution of our sample by 
applying the relation linking IRAC colors to redshifts \citep[see][]{Rettura2014, Martinez-Manso2015}. This IRAC color-redshift relation was derived using the multicolor and spectroscopic COSMOS/Ultra-Vista datasets \citep{Sanders2007}. 
Although the $z_p$--$[3.6]$-$[4.5]$ plane is
rather randomly populated, this approach works because for each color it is  possible to distinguish a most probable redshift
$\bar{z}_p$ and a typical error $\Delta z = 0.2-0.3$ for individual sources.  
We assigned a photometric redshift distribution for each cluster in our sample by taking the median 
redshift of the ten brightest (at $4.5\mu$m) cluster members (see prescription in \cite{Rettura2014}), an approach that 
minimizes contamination from faint sources which are more subject to photometric errors. The resulting
photometric redshift distribution is shown in Fig.~\ref{fig:redshift_distribution}.
The redshift distribution peaks at $z_p\!\simeq\!1.6$, with an extended tail at higher 
redshifts. We emphasize that this result must be taken with great caution as our photometric redshift distribution 
relies only on two-band photometry and results in a uncertaintly in the redshift, $\Delta z_p\!\sim\!0.7$ for the candidates. This large uncertainty is mainly due to the redshift-color relation no longer being valid for $z_p\!>\!1.7$ where the relation reaches a plateau.
However, we can still conclude that half of our overdensities have $z_p>1.7$, which is compatible with the redshifts we derive from the SPIRE photometry.

\subsection{Color-magnitude diagrams}
\label{subsect:CMD}

\begin{figure*}[!ht]
\begin{center}
\includegraphics[width=1.00\textwidth]{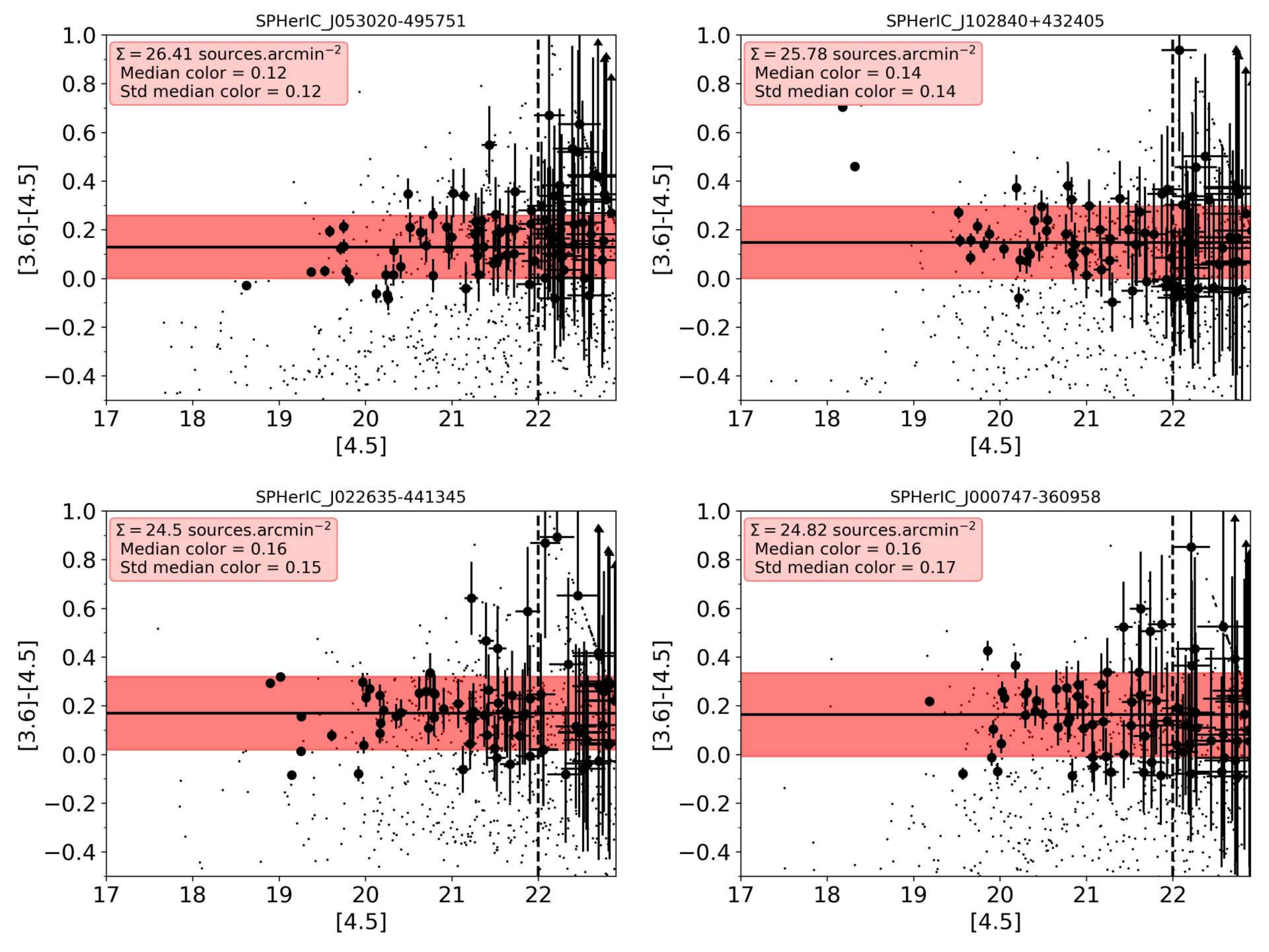}
\caption{For four representative clusters in our sample, $[3.6]-[4.5]$ color vs $[4.5]$ magnitude diagrams of red IRAC sources located within $1\arcmin$ of the BRCG position are shown (large black circles).  We represent all sources within the field for comparison (small black dots). Arrows indicate a non-detections at 3.6$\mu$m.  The median color for sources with $[4.5]<22$ is indicated by the solid horizontal and vertical dashed black lines, respectively. The $\pm 1 \sigma$ range in the mean color is indicated by the red shaded area. The values of the average surface density ($\Sigma$), mean color, and its standard deviation are given in each legend for each source.}
\label{fig:two_CMD}
\end{center}
\end{figure*}

As discussed in for example, \cite{Muzzin2013}, galaxies in high-$z$ clusters exhibit colors in a tight 
sequence which has been dubbed the ``Stellar Bump Sequence'' (SBS). The mean color of the galaxies that lie on the sequence can be used as a 
possible indicator of the redshift of overdense regions. For each SPHerIC candidate, we consider the 
IRAC color of each red IRAC source within a radius of $1\arcmin$ of each BRCG as a function of their
$4.5\mu$m magnitude. We calculated the mean color and its standard deviation for sources with $[4.5]<22$.
Four examples of
color-magnitude diagrams are presented in Fig.~\ref{fig:two_CMD}.
We find that the SPHerIC candidates have median colors, $[3.6]-[4.5]\!\sim\!0.1$, and standard 
deviations in the color similar to the other known clusters \citep{Muzzin2013, Rettura2014}. 
While not conclusive, color-magnitude diagrams like the ones exhibited by some of the SPHerIC candidates are consistent with their interpretation being bona-fide (proto)clusters.

\begin{figure*}[!ht]
\begin{center}
\includegraphics[width=1.00\textwidth]{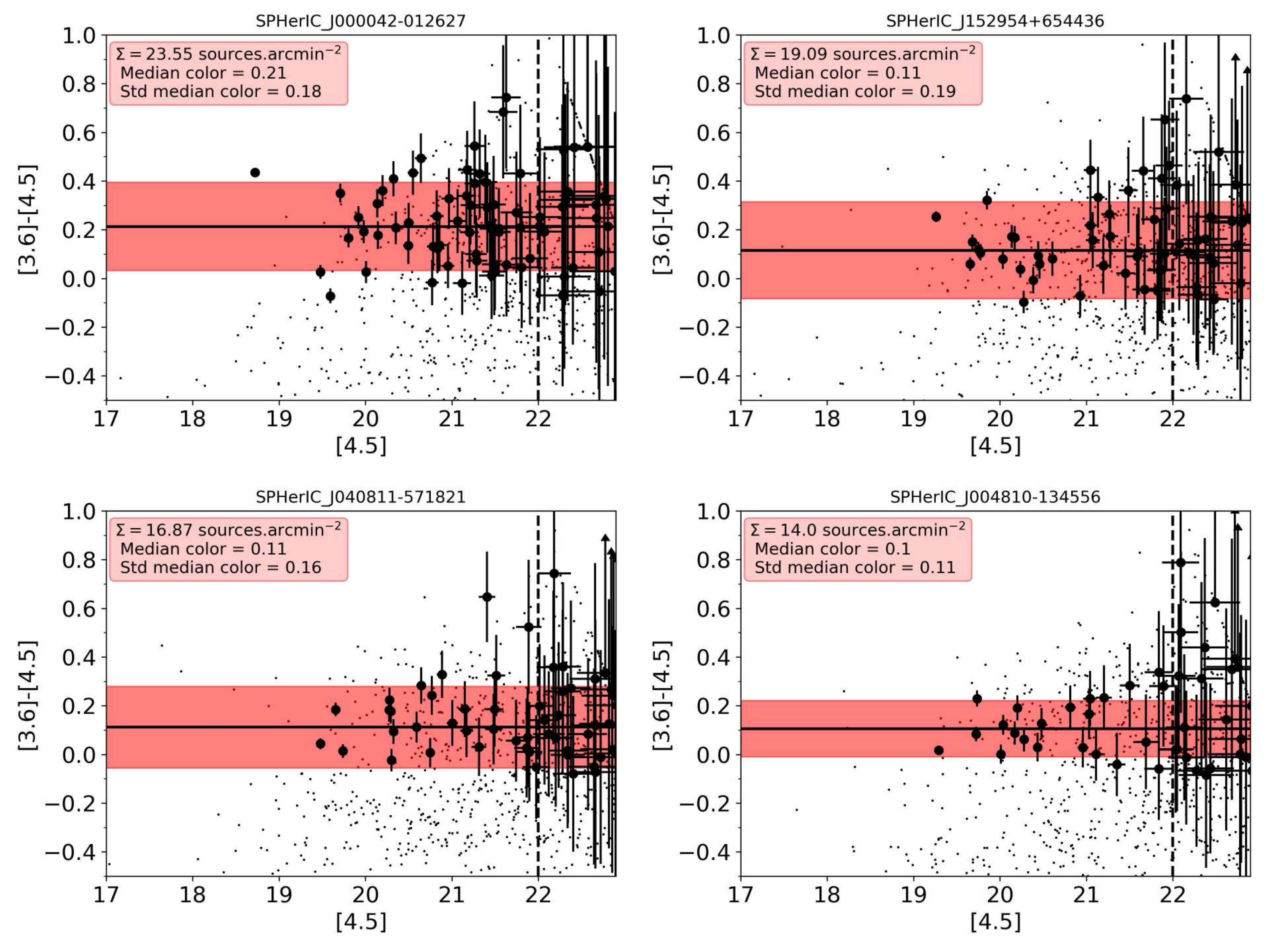}
\caption{Same legend as Fig. \ref{fig:two_CMD}. We show here SPHerIC\_J000042-012627, which is known to be a double structure at $z=1.7$ and $z=2$ \citep{Flores-Cacho2016}. We also show three examples of candidates exhibiting lower surface densities. The vast majority of our candidates exhibit color-magnitude diagrams compatible with high-$z$ (proto)clusters, although they seem indistinguishable from chance alignments. }
\label{fig:two_CMD2}
\end{center}
\end{figure*}

\subsection{Density maps}
\label{subsect:AKDE}

We have been considering the BRCGs as the centers of the galaxy overdensities.  Here  we test this assumption by constructing surface density maps of our candidates. 
We used an IDL (Interactive data language) implementation of the Adaptative Kernel Density Estimator \citep[AKDE][]{Pisani1996, Valtchanov2013}. The method is well understood and has been compared to other density estimators \citep{Ferdosi2011}. It assumes that the density field evaluated at the position, $\vec{r}$, is represented by the sum of Gaussian kernels centered on each source with a width of which depends on the local density of sources. The net effect is that in regions of large densities, where high resolution is warranted by the large density of sources, the width of the smoothing kernel will be small.  By contrast, in low density regions the with of the kernel will be large. 

We started by constructing density maps of randomly selected subregions in the SpUDS data with the same $\sim$5\arcmin$\times$5\arcmin size as our mosaics. These randomly selected regions will serve as our control fields. Within each control field, we estimated the surface density distribution of red IRAC sources and estimate a mean and standard deviation of the counts as we did for the IRAC images of our sample. We then constructed the significance of the overdensities in the SPHerICs maps in comparison to this mean density and its standard deviation. Examples of the resulting significance estimates for the SPHerIC fields are shown in Fig.~\ref{fig:top12_dens_maps_1} - \ref{fig:6_dens_maps_1}.

We first note that the SPherIC fields are all overdense on the scales of the individual IRAC mosaics, with all the fields in our sample having a mean density above the mean density derived from SpUDS density maps.  The total area covered by regions that are $3\sigma$ overdense ($A_{\sigma}$) are on average more than three times larger in the SPHerIC fields than in randomly sampled SpUDS regions. 
Specifically, out of 300 randomly selected fields within the total SpUDS imaging area, only 10 exhibit densities (within the uncertainties), similar to those in the SPHerIC survey. Within the SPHerIC survey fields, only ten of the 82 have surface densities of red sources that are consistent with the typical random fields from SpUDS. A Kolmogorov-Smirnov test indicates that the distributions of $A_{\sigma}$ in SpUDS and SPHerIC do not come from the same probability density function. This result implies that, statistically, the \Planck\ sources are overdense with red IRAC sources compared to random fields.

Next we wanted to test whether the BRCG is a reliable tracer of red IRAC source overdensities.
In total, the BRCG traces an overdensity of red IRAC sources in about 60\% our sample. In terms of angular size, the BRCG is located within 1\arcmin\ of the peak of the density map in 27\% of cases. In 33\% of cases, the BRCG traces not the maximum overdensity but a comparable one (within 30\%). Finally, we note that in 21\% of the fields, a red SPIRE source, but not the BRCG itself, is within 1\arcmin of the peak overdensity. So occasionally, overdensities are traced by red SPIRE sources which are not the brightest one in the field.

We show some examples where the BRCG does not trace the largest overdensity in the field in Fig.~\ref{fig:6_dens_maps_1}.  Thus, even when the BRCG does not lie in the most significant overdensity in the field, there are significant overdensities and some of those are associated with red SPIRE sources.  This suggests again that the \Planck\ selection does discover significant overdensities and accompanying SPIRE detections in these fields are meaningful signposts of overdensities (even if it is not always the brightest of the SPIRE sources that is directly located within an overdensity). 
The fact that vigorous star-formation seems to be associated with the densest environments at these redshifts is reminiscent of known high-$z$ (proto)structures \citep{Tran2010, Santos2015, Wang2016, Oteo2018}. \citet{Oteo2018} is particularly noteworthy as the overdensity they presented was discovered by searching for a galaxy overdensity around the reddest SPIRE source in H-ATLAS, a selection similar to ours. Perhaps this result and our results are consistent with a reversal of the SFR-$\rho$ relation, with enhanced, rapid star formation episodes occur in the most massive dark matter halos at high redshift \citep{Chiang2017}.

Finally, we note that the SPHerIC candidates have a single clear density peak in about $30\%$ of cases. Many of the fields have elongated overdense structures, and/or multiple overdense peaks of comparable significance. Even if projection effects play a role, such density distributions are expected for protoclusters where many substructures later merge to form a bona-fide cluster. For example, one protocluster candidate in our sample, SPHerIC\_J000042-012627, shows evidence for an alignment of two structures at $z=1.7$ and $z=2.0$ \citep{Flores-Cacho2016}.  We  note that the crude color-selection used here may miss out on parts of the  protocluster structure \citep[see e.g.,][]{Hatch2016}, thereby rendering the interpretation of a clean detection of a single peak inconclusive.  

\subsection{Halo masses of the SPHerIC candidates}\label{halomasses}

Precise estimates of the halo masses of the SPHerIC candidates are not yet possible to make without additional photometric and spectroscopic data.  However, we can obtain approximate estimates of the total mass of our candidates using a \Spitzer\ richness-mass relation calibrated using known clusters up to z$\sim$2 \citep{Rettura2017}. We note that this relation is calibrated using a sample of virialized clusters over the range $0.4<z<2$, while we estimate that $\ga$25\% of the SPHerIC cluster candidates lie at higher redshifts.However, at the redshift range of our sample, the large error in the individual photometric redshift estimates only marginally affect our results.  Thus, applying this relationship, we find that our sample has a mean mass of $\log <M_{500}/ M_{\odot}> = 14.3\pm0.2$ (1-$\sigma$). The uncertainty in the individual mass estimates are on the order $\sim0.3$ dex. We caution that our candidates are not expected to be virialized, and that the mass-richness relation could thus systematically overestimate their masses.  Keeping this caveat in mind, our mean mass estimate indicates that the SPHerIC (proto)cluster candidates are among the most massive objects at $1.3<z<3.0$.  Although crude, our results imply that our candidates are the likely progenitors of the most massive clusters observed locally. 

\section{Conclusions}\label{sect:conclusions}

In this paper we have presented a \Spitzer/IRAC 3.6 and 4.5$\mu$m imaging study of a sample of 82 high redshift galaxy (proto)cluster candidates, selected on the basis of their integrated star-formation rate 
using data from \Planck\ and \Herschel.  
Our \Spitzer\ imaging sample was selected in a two step process. The first step was a \Planck\ color selection that directly targets star-formation in dense environments at high redshift (1.5$\la z_p \la 3$), resulting in 2151 candidates on the cleanest 26\% of the all sky map \citep{Planck_XXXIX}. The analysis of the \Herschel/SPIRE data of a subsample containing 228 sources reveals an exceptional population of clustered FIR red sources with photometric redshifts, $z_{phot} \sim 2$, with high star-formation rates, $\sim750~M_{\odot}$ yr$^{-1}$ \citep{Planck_XXVII}. We imaged 82 of these fields with \Spitzer/IRAC around their brightest SPIRE red galaxy associated with each \Planck\ source and revealing the presence of large galaxy overdensities.

Our main conclusions from this analysis are as follow. 

We photometrically identified candidate cluster members on the basis of their IRAC colors, $[3.6]-[4.5]>-0.1$, in a $1\arcmin$ radius around the brightest red SPIRE source, the BRCG, in each of our fields.
We did this to test the hypothesis that these BRCGs are beacons tracing some of the most overdense regions in the Universe.

Our analysis of the surface density of red IRAC sources within 1 arcmin of the BRCGs implies that they are highly significant overdensities. Approximately 3/4 of our sample have overdensities that are $3 \sigma$ above the mean surface density of red IRAC sources of the field. We find our candidates have similar or higher overdensities of red IRAC selected sources compared other statistical \Spitzer\ cluster surveys (e.g., CARLA).

Our candidate (proto)clusters exhibit radial surface density profiles that imply indeed BRCGs are signposts of large galaxy overdensities. This new result is important as protocluster searches could be envisioned using data in the \Herschel\ archive or through exploiting future FIR survey missions.  Our candidates also exhibit tight color-magnitude relations in the $[3.6]-[4.5]$ vs. $[4.5]$ plane, with color distribution similar to limited number of other confirmed high-$z$ clusters.

Our clusters have mean estimated masses of $\log \langle M_{500}/M_{\odot} \rangle = 14.3 \pm 0.2$. Although these are likely over-estimates, they suggest our sample is comprised of progenitors of the most massive local clusters.

In conclusion, our study finds promising (proto)cluster candidates in their early mass assembly phase. However, spectroscopic data is needed to confirm cluster galaxy membership and to measure the total rate of star-formation within these overdensities. To this end, we have already started a large multiwavelength follow-up campaign of the SPHerIC sample which will enable us to obtain a deeper understanding of the nature of these structures and of the formation and evolution of massive clusters in the Universe.

\bibliographystyle{aa}
\bibliography{bib_test}

\begin{acknowledgements}
We thank Ranga Chary for his help, advice, and insights during the course of this study. This work is based in part on observations made with the Spitzer Space Telescope, which is operated by the Jet Propulsion Laboratory, California Institute of Technology under a contract with NASA. TheHerschel spacecraft was designed, built, tested, and launched under a contract to ESA managed by the Herschel/Planck Project team by an industrial consortium under the overall responsibility of the prime contractor Thales Alenia Space (Cannes), and including Astrium (Friedrichshafen) responsible for the payload module and for system testing at spacecraft level, Thales Alenia Space (Turin) responsible for the service module, and Astrium (Toulouse) responsible for the telescope, with in excess of a hundred subcontractors. The development of Planck has been supported by: ESA; CNES and CNRS/INSU-IN2P3-INP (France); ASI, CNR, and INAF (Italy); NASA and DoE (USA); STFC and UKSA (UK); CSIC, MICINN, JA, and RES (Spain); Tekes, AoF, and CSC (Finland); DLR and MPG (Germany); CSA (Canada); DTU Space (Denmark); SER/SSO (Switzerland); RCN (Norway); SFI (Ireland); FCT/MCTES (Portugal); and PRACE (EU). A description of the Planck Collaboration and a list of its members, including the technical or scientific activities in which they have been involved, can be found at \url{http:// www.sciops.esa.int/index.php?project=planck&page=Planck_Collaboration}. We acknowledge the support the PNCG (Programme National de Cosmologie et Galaxies). CM acknowledges the support provided by FONDECYT postdoctoral research grant no. 3170774. 
\end{acknowledgements}


\appendix
\section{Aperture corrections}
\label{app:aper_cor}
We under-estimate the fluxes due to using fixed apertures for the photometry, so we needed to perform aperture corrections. Aperture corrections are derived using \Spitzer/IRAC PSF files and by considering the enclosed flux in an aperture of growing radius $r$ (from $2\arcsec$ to $24\arcsec$). We took the flux in an aperture of diameter $24\arcsec$ to equal to the total flux and normalize the photometry by that value. Then the aperture correction, $\alpha(r)$, is then, 

\begin{equation}
\alpha(r) = \frac{f(r=24")}{f(r)} 
\end{equation}

The resulting aperture corrections are listed in Tables~\ref{tab:aper_cor_ch1} and A.2  for chs1 and 2 respectively. We find good agreement with our estimates and those from
the literature \citep{Barmby2008,Ashby2009}.

\begin{table}[!ht]
\centering
\begin{tabular}{lccc}
\hline 
\hline 
Aperture radius $r$ &$ 1.5"$ & $2.0"$ & $3.0"$  \\ 
\hline 
$\alpha(r)$ & 1.73 & 1.37 & 1.15 \\ 
$\alpha(r)$ \citep{Barmby2008} & 1.75 & 1.38 & 1.16 \\ 
$\alpha(r)$\citep{Ashby2009} & 1.85 & 1.42 & 1.17 \\ 
\hline 
\end{tabular} 
\caption{\label{tab:aper_cor_ch1} Aperture corrections (for fluxes in $\mu$Jy) for ch1 }
\end{table}

\begin{table}[!ht]
\centering
\begin{tabular}{lccc}
\hline 
\hline 
Aperture radius $r$ & $1.5"$ & $2.0"$ & $3.0"$  \\ 
\hline 
$\alpha(r)$  & 1.75 & 1.39 & 1.15 \\ 
$\alpha(r)$ \citep{Barmby2008} & 1.77 & 1.35 & 1.16 \\ 
$\alpha(r)$ \citep{Ashby2009} & 1.89 & 1.45 & 1.17 \\ 
\hline 
\end{tabular} 
\caption{\label{tab:aper_cor_ch2} Aperture corrections (for fluxes in $\mu$Jy) for ch2 }
\end{table}

\newpage

\section{Density maps}
We show here the surface density maps of red IRAC sources introduced in Section. \ref{subsect:AKDE}.
\begin{figure*}[!ht]
\begin{center}
\includegraphics[width=0.8\textwidth]{{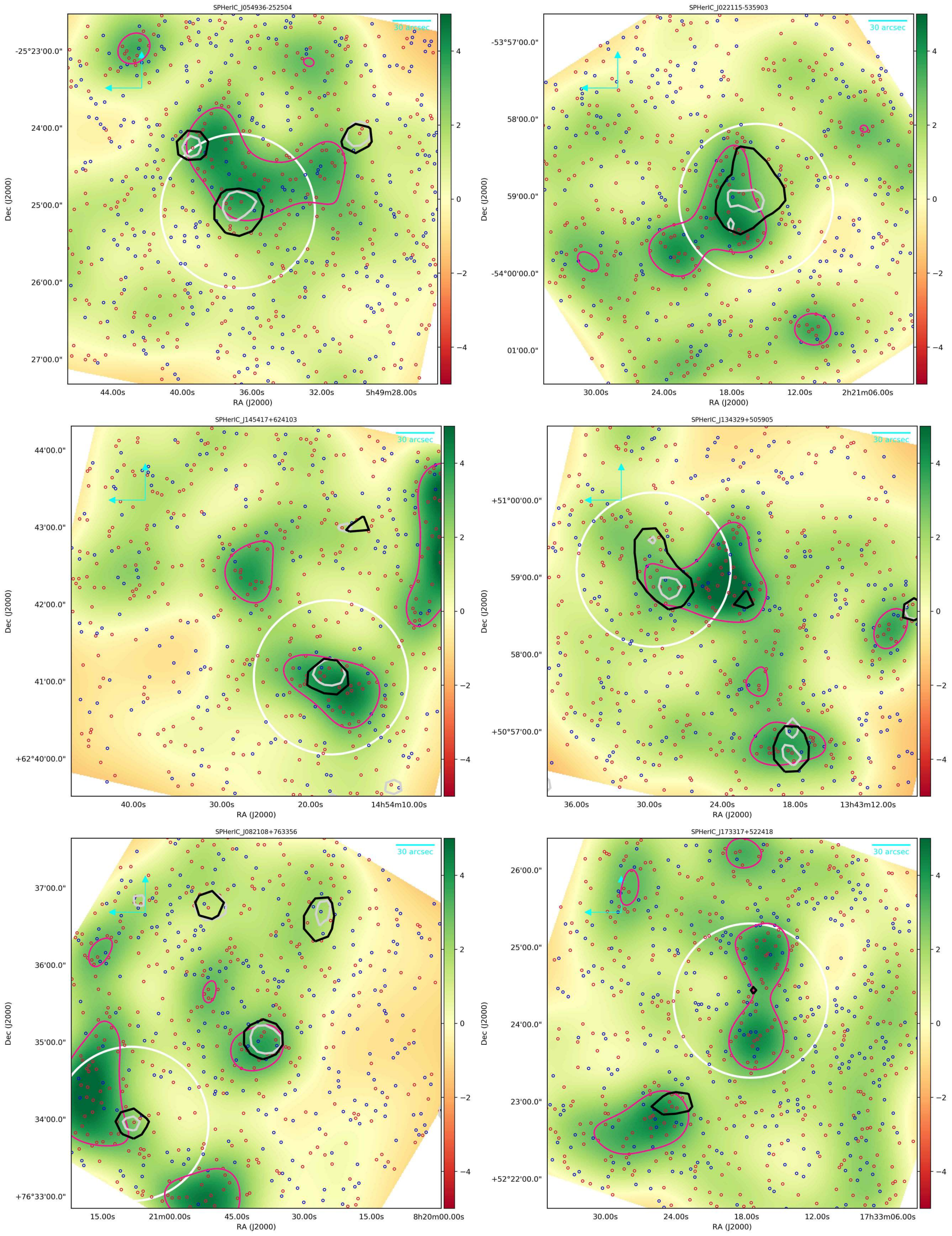}}
\caption{Overdensity significance maps of the SPHerIC candidates which are 4$\sigma$ overdense within a radius of 1\arcmin. Each image is 
$\sim$4.8$\arcmin\ \times$4.8$\arcmin$. IRAC sources with $[3.6]$--$[4.5]$ $>(<) -0.1$ are indicated by red (blue) circles. Contours at 3$\sigma$ significance of the overdensities are shown in pink. Herschel emission $3\sigma$ isocontours at 350 and 250$\mu$m are shown as black and gray lines respectively. The large white circle has a radius of $1\arcmin$ and is centered on the peak of emission at $350\mu$m from SPIRE, as already shown in Fig.~\ref{fig:top12_1}.}
\label{fig:top12_dens_maps_1}
\end{center}
\end{figure*}

\begin{figure*}[!ht]
\begin{center}
\includegraphics[width=0.8\textwidth]{{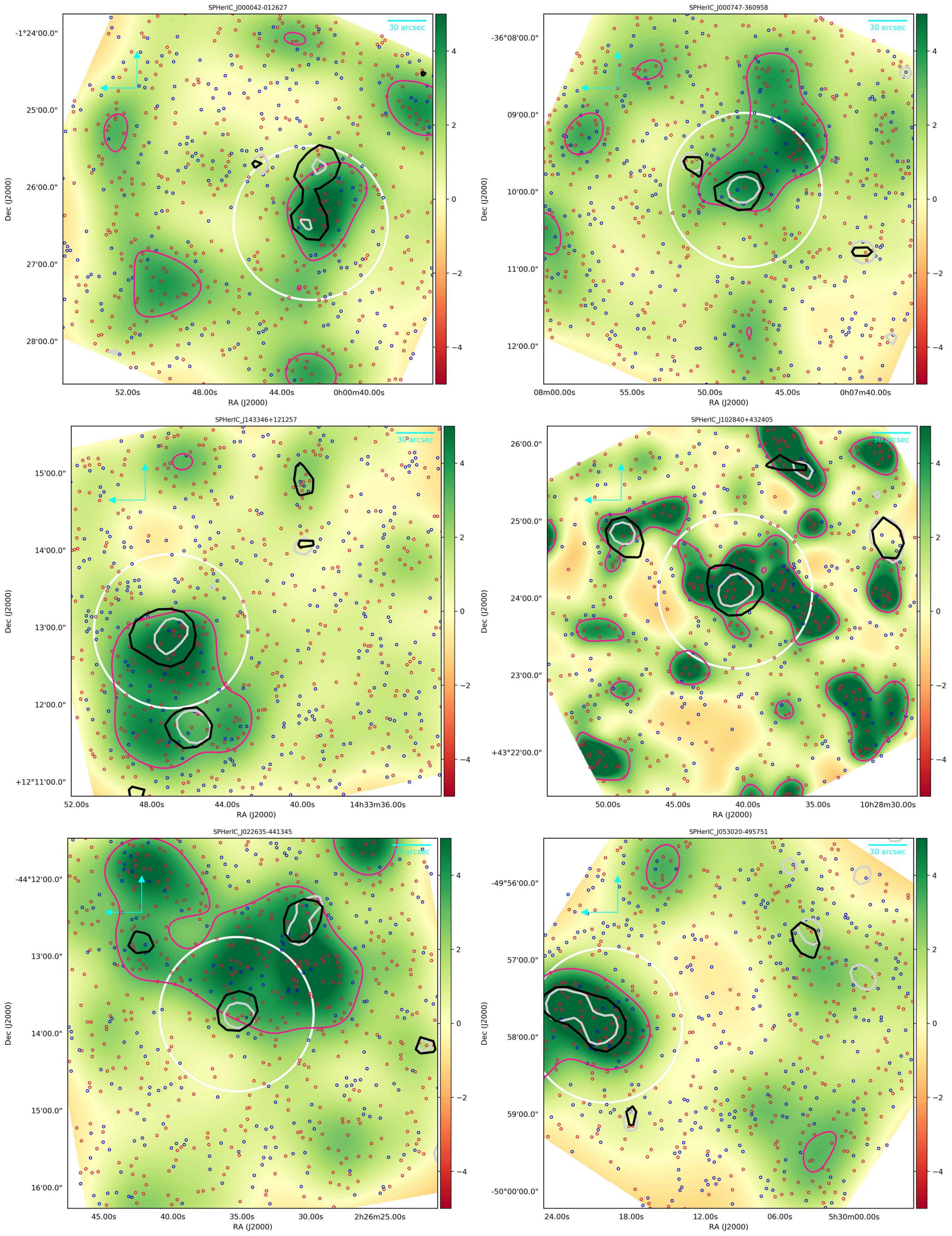}}
\caption{Continued from Fig.~\ref{fig:top12_dens_maps_1}.}
\label{fig:top12_dens_maps_2}
\end{center}
\end{figure*}

%

\begin{figure*}[!ht]
\begin{center}
\includegraphics[width=0.8\textwidth]{{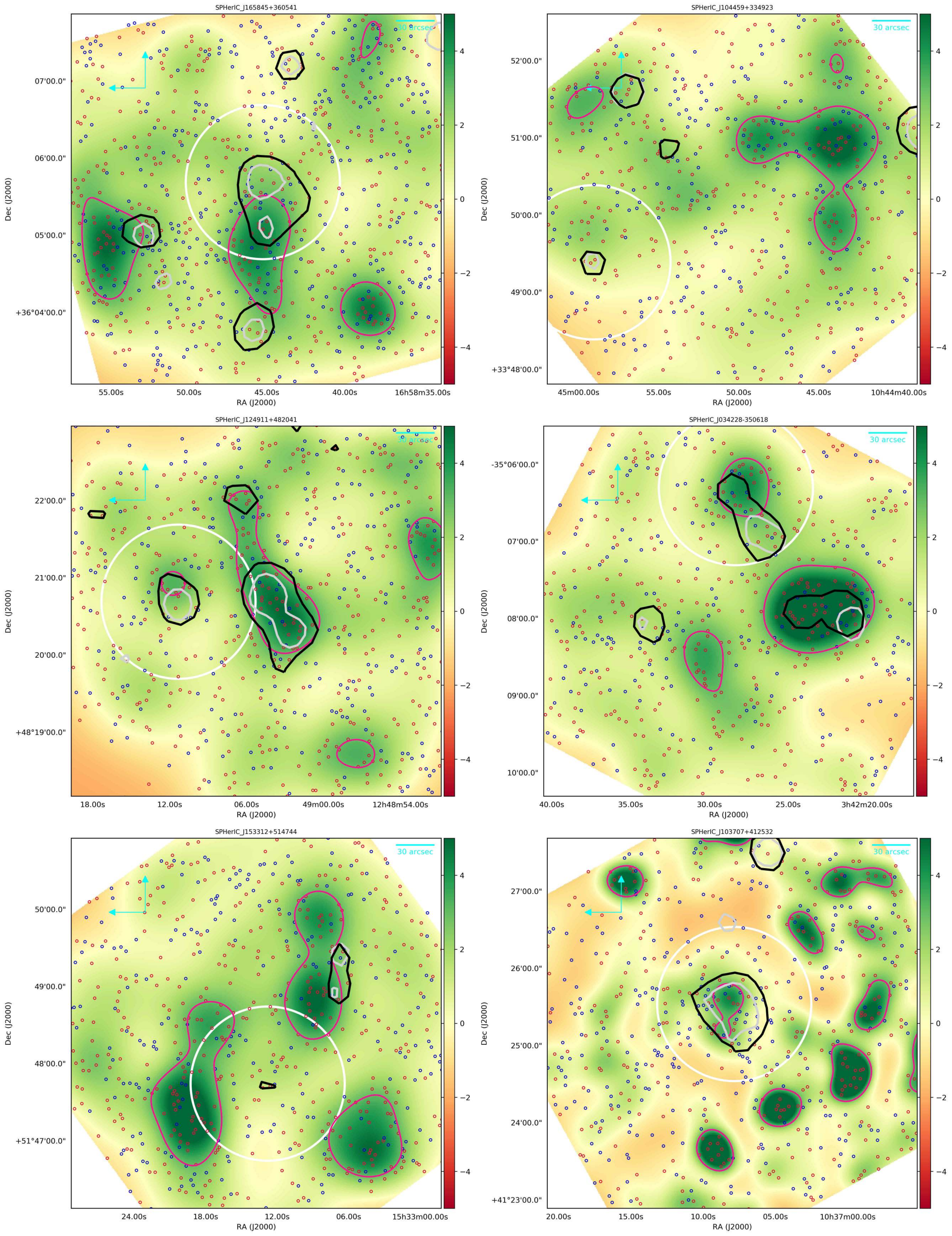}}
\caption{Overdensity significance maps of SPHerIC candidates where a significant overdensity is detected but it is not within 1 arcmin of the BRCG or an even more significant overdensity is present in the field outside of 1 arcmin from the BCCG. Symbols are the same as in Fig. \ref{fig:top12_dens_maps_1}.}
\label{fig:6_dens_maps_1}
\end{center}
\end{figure*}

\end{document}